\catcode`\@=11                                   % To make protected \def's

%************************************************************
%*
%*            Font set-up
%*
%************************************************************

%************** 5-point fonts *******************************

\font\fiverm=cmr5                         % roman
\font\fivemi=cmmi5                        % math italic
\font\fivesy=cmsy5                        % math symbols
\font\fivebf=cmbx5                        % bold face

\skewchar\fivemi='177
\skewchar\fivesy='60

%************** 6-point fonts *******************************

\font\sixrm=cmr6                          % roman
\font\sixi=cmmi6                          % math italic
\font\sixsy=cmsy6                         % math symbols
\font\sixbf=cmbx6                         % bold face

\skewchar\sixi='177
\skewchar\sixsy='60

%************** 7-point fonts *******************************

\font\sevenrm=cmr7                        % roman
\font\seveni=cmmi7                        % math italic
\font\sevensy=cmsy7                       % math symbols
\font\sevenit=cmti7                       % italic
\font\sevenbf=cmbx7                       % bold face

\skewchar\seveni='177
\skewchar\sevensy='60

%************** 8-point fonts *******************************

\font\eightrm=cmr8                        % roman
\font\eighti=cmmi8                        % math italic
\font\eightsy=cmsy8                       % math symbols
\font\eightit=cmti8                       % italic
                       % slanted
\font\eightbf=cmbx8                       % bold face
                       % typewriter
                       % sans serif

\skewchar\eighti='177
\skewchar\eightsy='60

%************** 9-point fonts *******************************

\font\ninei=cmmi9
\font\ninesy=cmsy9

\skewchar\ninei='177
\skewchar\ninesy='60

%************** 10-point fonts ******************************

\font\tenrm=cmr10                         % roman
\font\teni=cmmi10                         % math italic
\font\tensy=cmsy10                        % math symbols
\font\tenex=cmex10                        % math extension
\font\tenit=cmti10                        % italic
\font\tensl=cmsl10                        % slanted
\font\tenbf=cmbx10                        % bold face
\font\tentt=cmtt10                        % typewriter
\font\tenss=cmss10                        % sans serif
\font\tensc=cmcsc10                       % small caps
\font\tenbi=cmmib10                       % bold math

\skewchar\teni='177
\skewchar\tenbi='177
\skewchar\tensy='60

\def\tenpoint{\ifmmode\err@badsizechange\else
	\textfont0=\tenrm \scriptfont0=\sevenrm \scriptscriptfont0=\fiverm
	\textfont1=\teni  \scriptfont1=\seveni  \scriptscriptfont1=\fivemi
	\textfont2=\tensy \scriptfont2=\sevensy \scriptscriptfont2=\fivesy
	\textfont3=\tenex \scriptfont3=\tenex   \scriptscriptfont3=\tenex
	\textfont4=\tenit \scriptfont4=\sevenit \scriptscriptfont4=\sevenit
	\textfont5=\tensl
	\textfont6=\tenbf \scriptfont6=\sevenbf \scriptscriptfont6=\fivebf
	\textfont7=\tentt
	\textfont8=\tenbi \scriptfont8=\seveni  \scriptscriptfont8=\fivemi
	\def\rm{\tenrm\fam=0 }%
	\def\it{\tenit\fam=4 }%
	\def\sl{\tensl\fam=5 }%
	\def\bf{\tenbf\fam=6 }%
	\def\tt{\tentt\fam=7 }%
	\def\ss{\tenss}%
	\def\sc{\tensc}%
	\def\bmit{\fam=8 }%
	\rm\setparameters\setbaselines\fi}

%************** 12-point fonts ******************************

\font\twelverm=cmr12                      % roman
\font\twelvei=cmmi12                      % math italic
\font\twelvesy=cmsy10       scaled\magstep1             % math symbols
\font\twelveex=cmex10       scaled\magstep1             % math extension
\font\twelveit=cmti12                            % italic
\font\twelvesl=cmsl12                            % slanted
\font\twelvebf=cmbx12                            % bold face
\font\twelvett=cmtt12                            % typewriter
\font\twelvess=cmss12                            % sans serif
\font\twelvesc=cmcsc10      scaled\magstep1             % small caps
\font\twelvebi=cmmib10      scaled\magstep1             % bold math

\skewchar\twelvei='177
\skewchar\twelvebi='177
\skewchar\twelvesy='60

\def\twelvepoint{\ifmmode\err@badsizechange\else
	\textfont0=\twelverm \scriptfont0=\eightrm \scriptscriptfont0=\sixrm
	\textfont1=\twelvei  \scriptfont1=\eighti  \scriptscriptfont1=\sixi
	\textfont2=\twelvesy \scriptfont2=\eightsy \scriptscriptfont2=\sixsy
	\textfont3=\twelveex \scriptfont3=\tenex   \scriptscriptfont3=\tenex
	\textfont4=\twelveit \scriptfont4=\eightit \scriptscriptfont4=\sevenit
	\textfont5=\twelvesl
	\textfont6=\twelvebf \scriptfont6=\eightbf \scriptscriptfont6=\sixbf
	\textfont7=\twelvett
	\textfont8=\twelvebi \scriptfont8=\eighti  \scriptscriptfont8=\sixi
	\def\rm{\twelverm\fam=0 }%
	\def\it{\twelveit\fam=4 }%
	\def\sl{\twelvesl\fam=5 }%
	\def\bf{\twelvebf\fam=6 }%
	\def\tt{\twelvett\fam=7 }%
	\def\ss{\twelvess}%
	\def\sc{\twelvesc}%
	\def\bmit{\fam=8 }%
	\rm\setparameters\setbaselines\fi}

%************** 14-point fonts ******************************

\font\fourteenrm=cmr12      scaled\magstep1             % roman
\font\fourteeni=cmmi12      scaled\magstep1             % math italic
\font\fourteensy=cmsy10     scaled\magstep2             % math symbols
\font\fourteenex=cmex10     scaled\magstep2             % math extension
\font\fourteenit=cmti12     scaled\magstep1             % italic
\font\fourteensl=cmsl12     scaled\magstep1             % slanted
\font\fourteenbf=cmbx12     scaled\magstep1             % bold face
\font\fourteentt=cmtt12     scaled\magstep1             % typewriter
\font\fourteenss=cmss12     scaled\magstep1             % sans serif
\font\fourteensc=cmcsc10 scaled\magstep2  % small caps
\font\fourteenbi=cmmib10 scaled\magstep2  % bold math

\skewchar\fourteeni='177
\skewchar\fourteenbi='177
\skewchar\fourteensy='60

\def\fourteenpoint{\ifmmode\err@badsizechange\else
	\textfont0=\fourteenrm \scriptfont0=\tenrm \scriptscriptfont0=\sevenrm
	\textfont1=\fourteeni  \scriptfont1=\teni  \scriptscriptfont1=\seveni
	\textfont2=\fourteensy \scriptfont2=\tensy \scriptscriptfont2=\sevensy
	\textfont3=\fourteenex \scriptfont3=\tenex \scriptscriptfont3=\tenex
	\textfont4=\fourteenit \scriptfont4=\tenit \scriptscriptfont4=\sevenit
	\textfont5=\fourteensl
	\textfont6=\fourteenbf \scriptfont6=\tenbf \scriptscriptfont6=\sevenbf
	\textfont7=\fourteentt
	\textfont8=\fourteenbi \scriptfont8=\tenbi \scriptscriptfont8=\seveni
	\def\rm{\fourteenrm\fam=0 }%
	\def\it{\fourteenit\fam=4 }%
	\def\sl{\fourteensl\fam=5 }%
	\def\bf{\fourteenbf\fam=6 }%
	\def\tt{\fourteentt\fam=7}%
	\def\ss{\fourteenss}%
	\def\sc{\fourteensc}%
	\def\bmit{\fam=8 }%
	\rm\setparameters\setbaselines\fi}

%************** Miscellaneous big fonts *********************

\font\seventeenrm=cmr10 scaled\magstep3          % roman
             % bold face

%************************************************************
%*
%*            Parameter initialization
%*
%************************************************************

\newdimen\rp@
\newcount\@basestretchnum
\newskip\@baseskip
\newskip\headskip
\newskip\footskip

% Routine to set page parameters

\def\setparameters{\rp@=.1em
	\headskip=24\rp@
	\footskip=\headskip
	\delimitershortfall=5\rp@
	\nulldelimiterspace=1.2\rp@
	\scriptspace=0.5\rp@
	\abovedisplayskip=10\rp@ plus3\rp@ minus5\rp@
	\belowdisplayskip=10\rp@ plus3\rp@ minus5\rp@
	\abovedisplayshortskip=5\rp@ plus2\rp@ minus4\rp@
	\belowdisplayshortskip=10\rp@ plus3\rp@ minus5\rp@
	\normallineskip=\rp@
	\lineskip=\normallineskip
	\normallineskiplimit=0pt
	\lineskiplimit=\normallineskiplimit
	\jot=3\rp@
	\setbox0=\hbox{\the\textfont3 B}\p@renwd=\wd0
	\skip\footins=12\rp@ plus3\rp@ minus3\rp@
	\skip\topins=0pt plus0pt minus0pt}

% Special routine to scale \baselineskip

\def\setbaselines{\maxdepth=4\rp@\baselinestretch=\@basestretchnum}

% The \baselinestretch command

\def\baselinestretch{\afterassignment\@basestretch\@basestretchnum}
\def\@basestretch{%
	\@baseskip=12\rp@ \divide\@baseskip by1000
	\normalbaselineskip=\@basestretchnum\@baseskip
	\baselineskip=\normalbaselineskip
	\bigskipamount=\the\baselineskip
		plus.25\baselineskip minus.25\baselineskip
	\medskipamount=.5\baselineskip
		plus.125\baselineskip minus.125\baselineskip
	\smallskipamount=.25\baselineskip
		plus.0625\baselineskip minus.0625\baselineskip
	\setbox\strutbox=\hbox{\vrule height.708\baselineskip
		depth.292\baselineskip width0pt }}

%************************************************************
%*
%*            Modifications to PLAIN.TEX
%*
%************************************************************

% Modifications to PLAIN routines to handle scaling of page parameters

\def\makeheadline{\vbox to0pt{\baselinestretch=1000
	\vskip-\headskip \vskip1.5pt
	\line{\vbox to\ht\strutbox{}\the\headline}\vss}\nointerlineskip}

\def\makefootline{\baselineskip=\footskip\line{\the\footline}}

\def\big#1{{\hbox{$\left#1\vbox to8.5\rp@ {}\right.\n@space$}}}
\def\Big#1{{\hbox{$\left#1\vbox to11.5\rp@ {}\right.\n@space$}}}
\def\bigg#1{{\hbox{$\left#1\vbox to14.5\rp@ {}\right.\n@space$}}}
\def\Bigg#1{{\hbox{$\left#1\vbox to17.5\rp@ {}\right.\n@space$}}}

% Modifications to PLAIN to handle bold math

\mathchardef\alpha="710B
\mathchardef\beta="710C
\mathchardef\gamma="710D
\mathchardef\delta="710E
\mathchardef\epsilon="710F
\mathchardef\zeta="7110
\mathchardef\eta="7111
\mathchardef\theta="7112
\mathchardef\iota="7113
\mathchardef\kappa="7114
\mathchardef\lambda="7115
\mathchardef\mu="7116
\mathchardef\nu="7117
\mathchardef\xi="7118
\mathchardef\pi="7119
\mathchardef\rho="711A
\mathchardef\sigma="711B
\mathchardef\tau="711C
\mathchardef\upsilon="711D
\mathchardef\phi="711E
\mathchardef\chi="711F
\mathchardef\psi="7120
\mathchardef\omega="7121
\mathchardef\varepsilon="7122
\mathchardef\vartheta="7123
\mathchardef\varpi="7124
\mathchardef\varrho="7125
\mathchardef\varsigma="7126
\mathchardef\varphi="7127
\mathchardef\imath="717B
\mathchardef\jmath="717C
\mathchardef\ell="7160
\mathchardef\wp="717D
\mathchardef\partial="7140
\mathchardef\flat="715B
\mathchardef\natural="715C
\mathchardef\sharp="715D

%************************************************************
%*
%*            Initialization
%*
%************************************************************

\def\err@badsizechange{%
	\immediate\write16{--> Size change not allowed in math mode, ignored}}

\baselinestretch=1000
\tenpoint

\catcode`\@=12                                   % Restore @ sign
% Routine to guarantee that this file is input only once
\catcode`\@=11
\expandafter\ifx\csname @iasmacros\endcsname\relax
	\global\let\@iasmacros=\par
\else  \immediate\write16{}
	\immediate\write16{Warning:}
	\immediate\write16{You have tried to input iasmacros more than once.}
	\immediate\write16{}
	\endinput
\fi
\catcode`\@=12

% Set up font size commands and \baselinestretch command
%\input iasfonts

% Some alternative font names
\def\rmb{\seventeenrm}

% Simple spacing commands
\def\singlespace{\baselineskip=\normalbaselineskip}
\def\halfspace{\baselineskip=1.5\normalbaselineskip}
\def\doublespace{\baselineskip=2\normalbaselineskip}

% Macros for references and abstracts

\def\AB{\bigskip\parindent=40pt
	 \centerline{\bf ABSTRACT}\medskip\halfspace\narrower}
\def\AE{\bigskip\nonarrower\doublespace}
\def\nonarrower{\advance\leftskip by-\parindent
	\advance\rightskip by-\parindent}

% Useful commands

\def\boxit#1{\vbox{\hrule\hbox{\vrule\kern3pt
	\vbox{\kern3pt#1\kern3pt}\kern3pt\vrule}\hrule}}

% Special symbols
\def\hence{\leavevmode\hbox{\bf .\raise5.5pt\hbox{.}.} }

\def\dalemb#1#2{{\vbox{\hrule height.#2pt
	\hbox{\vrule width.#2pt height#1pt \kern#1pt \vrule width.#2pt}
	\hrule height.#2pt}}}
\def\gtorder{\mathrel{\raise.3ex\hbox{$>$}\mkern-14mu
	      \lower0.6ex\hbox{$\sim$}}}
\def\ltorder{\mathrel{\raise.3ex\hbox{$<$}\mkern-14mu
	      \lower0.6ex\hbox{$\sim$}}}

% For twoup output
\newdimen\fullhsize
\newbox\leftcolumn
\def\twoup{\hoffset=-.5in \voffset=-.25in
  \hsize=4.75in \fullhsize=10in \vsize=6.9in
  \def\fullline{\hbox to\fullhsize}
  \let\lr=L
  \output={\if L\lr
	 \global\setbox\leftcolumn=\columnbox\global\let\lr=R \advancepageno
      \else \doubleformat \global\let\lr=L\fi
    \ifnum\outputpenalty>-20000 \else\dosupereject\fi}
  \def\doubleformat{\shipout\vbox{
    \fullline{\box\leftcolumn\hfil\columnbox}\advancepageno}}
  \def\columnbox{\leftline{\vbox{\makeheadline\pagebody\makefootline}}}
  \tolerance=1000 }
\twelvepoint
\doublespace
\overfullrule=0pt
{\nopagenumbers{

\rightline{~~~April, 2007}
\bigskip\bigskip
\centerline{\rmb A density tensor hierarchy for open}
\centerline{\rmb system dynamics: retrieving the noise}
\medskip
\centerline{\it  Stephen L. Adler}
\centerline{\bf Institute for Advanced Study}
\centerline{\bf Princeton, NJ 08540}
\medskip
\bigskip\bigskip
\leftline{\it Send correspondence to:}
\medskip
{\singlespace\leftline{Stephen L. Adler}
\leftline{Institute for Advanced Study}
\leftline{Einstein Drive, Princeton, NJ 08540}
\leftline{Phone 609-734-8051; FAX 609-924-8399; 
email adler@ias.edu}}
\bigskip\bigskip
}}
\vfill\eject
\pageno=2
\AB
We develop a density tensor hierarchy for open system dynamics, that 
recovers information about fluctuations (or ``noise'') lost           
in passing to the reduced density matrix.  For the case of fluctuations 
arising from a classical probability distribution, the hierarchy is formed
from expectations of products of pure state density matrix elements, and   
can be compactly summarized by a simple generating function.  
For the case of quantum fluctuations arising when a quantum system interacts 
with a quantum environment in an overall pure state, the corresponding 
hierarchy is defined as the environmental trace of products of system matrix 
elements of the full density matrix.  Whereas all members of the classical 
noise hierarchy are system observables, only the lowest member of the 
quantum noise hierarchy is directly experimentally measurable.  
The unit trace and idempotence  
properties of the pure state density matrix imply descent relations for 
the tensor hierarchies, that relate the order $n$ tensor, under contraction 
of appropriate pairs of tensor indices, to the order $n-1$ tensor.  
As examples to illustrate the classical probability distribution formalism, 
we consider 
a spatially isotropic ensemble of spin-1/2 pure states, a  quantum 
system evolving by an It\^o  stochastic Schr\"odinger equation, and a quantum 
system evolving by a jump process Schr\"odinger equation.  
As examples to illustrate the corresponding trace formalism in  
the quantum fluctuation case, we consider the tensor hierarchies for  
collisional Brownian motion of an infinite mass Brownian particle, and for  
the the weak coupling 
Born-Markov master equation. In different specializations, the latter gives 
the hierarchies generalizing the quantum optical master equation and 
the Caldeira--Leggett master equation.   As a further application of the 
density   
tensor, we contrast stochastic Schr\"odinger equations that reduce 
and that do not reduce the state vector, and discuss why a quantum system 
coupled to a quantum environment behaves like the latter.  
The descent relations for  
our various examples are checked in a series of Appendices.

\AE
\bigskip\bigskip
\vfill\eject
\pageno=3
\centerline{{\bf 1.~~Introduction}}
\bigskip
Increasing attention is being paid to the dynamics of open quantum systems, 
that is, to quantum systems acted on by an environment.  
Such systems are of interest for studies of dissipative 
phenomena, decoherence, backgrounds to quantum computers and to precision 
measurements, and theories of quantum measurement.  A principal tool in  
studying  open quantum systems is the reduced density matrix, obtained from 
the pure state density matrix by tracing over environment degrees of freedom, 
or in stochastic models where the environment is represented by a noise 
term in the Schr\"odinger equation, by averaging over the noise.   As is 
well-known, this transition from the pure state density matrix to the reduced 
density matrix is not one-to-one, since information about the total system 
is lost.  For example, in stochastic models, there is known to be a 
continuum of different unravelings, or pure state density matrix stochastic 
evolutions, that yield the same master equation for the reduced density 
matrix.  The question that we investigate here is the extent to which one 
can form objects that refer only to the basis vectors of the system 
Hilbert space, but that nonetheless recapture  information 
that is lost in passing to the reduced density matrix.  In  the first part 
of this  
paper (Sections 2 through 5), we discuss classical noise 
arising from fluctuations defined by classical probability 
distributions.  In the second part (Sections 6 through 9), we give an 
analogous discussion of quantum 
noise, which appears in the physically important  case of a quantum 
system coupled to a quantum environment in an overall pure state.    
We  also give an extension, making contact with the 
discussion of the first part, 
to the case in which the overall system is in a mixed state 
superposition of pure states.  The final section contains a discussion  
of quantum measurements that relates the material in the first 
and second parts.  

For the case of classical probability distributions, 
a relevant discussion  appears in Chapter 5 of 
the book {\it The 
Theory of Open Quantum Systems} by Breuer and Petruccione [1], following 
up on earlier papers by those authors [2], by Wiseman [3] and by 
M\o lmer, Castin, and Dalibard [4].  In simplified form,  
Breuer and Petruccione introduce  
an ensemble of pure state vectors $|\psi_{\alpha}\rangle$, each drawn from  
the same system Hilbert space ${\cal H}_{\cal S}$, 
with each vector assumed to occur in the ensemble 
with probability $w_{\alpha}$,  $\sum_{\alpha}w_{\alpha}=1$. 
Measurement of a general self-adjoint operator $R$ for a system prepared 
in $|\psi_{\alpha}\rangle$ typically gives a range of values, the 
mean of which 
given by $\langle \psi_{\alpha}| R | \psi_{\alpha} \rangle$. The mean or  
expectation over the ensemble of pure state vectors is then given by 
$$\sum_{\alpha} w_{\alpha} \langle \psi_{\alpha}| R | \psi_{\alpha} \rangle
={\rm Tr} \rho R~~~,\eqno(1a)$$
with $\rho$ the mixed state or reduced density matrix defined by 
$$\rho=\sum_{\alpha} w_{\alpha} |\psi_{\alpha} \rangle \langle \psi_{\alpha}|
~~~. \eqno(1b)$$
Breuer and Petruccione point out that there are three variances that are 
relevant.  The variance of measurements of $R$ over all pure states in the 
ensemble is given by        
$${\rm Var}(R)={\rm Tr} \rho (R- {\rm Tr} \rho R)^2 =
{\rm Tr} \rho R^2 - ({\rm Tr} \rho R)^2~~~.\eqno(2a)$$
This can be written as the sum of two non-negative terms, 
$${\rm Var}(R)={\rm Var}_1(R)+{\rm Var}_2(R)~~~,\eqno(2b)$$ 
with ${\rm Var}_1(R)$ 
the ensemble average of the variances of $R$ within each pure state 
of the ensemble, 
$${\rm Var}_1(R)=\sum_{\alpha} w_{\alpha} 
[\langle \psi_{\alpha}|R^2|\psi_{\alpha} \rangle  
-\langle \psi_{\alpha}|R|\psi_{\alpha} \rangle^2]~~~,\eqno(2c)$$ 
and with ${\rm Var}_2(R)$ the variance of the pure state means of 
$R$ over the ensemble, 
$${\rm Var}_2(R)=\sum_{\alpha} w_{\alpha} 
\langle \psi_{\alpha}| R | \psi_{\alpha} \rangle^2 
-[\sum_{\alpha} w_{\alpha} \langle \psi_{\alpha}| R | \psi_{\alpha} \rangle]^2
~~~.\eqno(2d)$$
Thus, ${\rm Var}_1(R)$ is an ensemble average of the quantum variances of 
$R$, while ${\rm Var}_2(R)$ is a measure of the spread of the average 
values of $R$ resulting from the statistical properties of the ensemble.  
As Breuer and Petruccione note, neither of the subsidiary variances 
${\rm Var}_{1,2}$ can be expressed as the density matrix expectation of 
some self-adjoint operator.  

Our aim in the first part  of this paper is to extend 
the formalism of ref [1] by utilizing  a density tensor hierarchy, which 
captures the statistical information that is lost in forming the reduced 
density matrix of Eq.~(1b).  
A density tensor, defined as an ensemble average of density matrices,  
was first introduced by Mielnik [5], and was applied to discussions 
of density functions on the space of quantum states and their 
application  
to thermalization of quantum systems by Brody and 
Hughston [6].  These papers, in addition to introducing the concept 
of a density tensor which is developed further here, also  
contain the important result 
that in the case of 
a continuum probability distribution, the density tensor 
hierarchy gives all of the information needed to reconstruct the 
probability function $w_{\alpha}$.  In particular,  the
variances ${\rm Var}_{1,2}$ for any 
observable, and more 
general statistical properties of the ensemble as well, can be expressed 
as contractions of density tensor matrix elements with appropriate matrix 
elements of the observable(s) of interest.  

The basic construction of the density tensor hierarchy corresponding to 
a classical probability distribution $\{w_{\alpha}\}$ is given in Sec. 2. 
Here  we generalize the reduced density matrix of Eq.~(1b) to 
a  density tensor, formed by taking a product of pure state density 
matrix elements, and averaging over the ensemble of pure states.   
When the $\psi_{\alpha}$ are independent of $\alpha$, 
this tensor reduces to an $n$-fold product of reduced density matrices,   
and so the difference between the density tensor and this product is a measure 
of the statistical fluctuations in the ensemble.  In the generic case of 
non-trivial dependence of $\psi_{\alpha}$ on $\alpha$, there are some 
general statements that can be made.  First of all, the order $n$  
density tensor is a symmetric tensor in its pair indices, 
and it can be considered   
as a matrix operator acting on the $n$-fold tensor product of the 
system Hilbert space ${\cal H}_{\cal S}$ with itself.   The symmetry of 
the density tensor allows construction of a generating function that 
on expansion gives the density tensors of all orders.  
Additionally, as a consequence of 
the unit trace and idempotence conditions obeyed by the pure state density 
matrix, the density tensor hierarchy satisfies a system of descent 
equations, relating the order $n$ tensor to the order $n-1$ tensor when 
any  row index is contracted with any column index.   
We show that the variances 
${\rm Var}_{1,2}$ defined by Breuer and Petruccione can be expressed in terms 
of appropriate contractions of density tensor elements with operator matrix 
elements.  

In subsequent sections  we develop some concrete applications of the 
general formalism for classical probability distributions.  In Sec. 3, we consider an isotropic ensemble of 
spin-1/2 pure state density matrices, construct the density tensors 
through order 3, verify the descent equations, and calculate the   
generating function. In Sec. 4 we apply 
the formalism to a quantum system evolving under the influence of noise 
as described by a stochastic Schr\"odinger equation, with the ensemble   
defined as the set of all histories of an initial quantum state under 
the influence of the noise.  Assuming white noise described by the 
It\^o calculus, we give the 
dynamics of the general density tensor 
in terms of the general unraveling of the Lindblad equation 
constructed by Wiseman and Di\'osi  [7], and show that the order two 
and higher   
density tensors distinguish between inequivalent unravelings that give  
the same reduced density matrix (i.e., the same order one density tensor).  
In Sec. 5 we develop an analogous formalism for the case of jump (piecewise 
deterministic process) unravelings of the Lindblad equation.  

We turn next to an analysis of a quantum system coupled to a quantum 
environment, rather than to an external classical noise source.  
Here, one 
is confronted with the problem of discussing the system dissipation associated 
with the system-environment interaction within a single overall pure state 
of system plus environment 
(or in a thermal state that is a weighted average of such pure states).  
Typically, in master equation derivations, the 
system-environment interaction\footnote{$^1$}
{What we call $H$ is usually  
denoted by $H_I$ in the open systems literature.  To avoid confusion, all 
other Hamiltonians will carry subscripts, e.g., $H_{\cal S}$ and $H_{\cal E}$ 
for the system and environment Hamiltonians, $H_{\rm TOT}$ for the total 
Hamiltonian, etc.}
$H$ has vanishing expectation 
in the environment, but its square $H^2$ does not have a vanishing 
expectation, because the environment is not in an eigenstate of $H$.  
The associated variance is then a measure of quantum fluctuations associated 
with the environment state, and is the source of quantum ``noise'' driving 
the system dissipation. Our aim in the second part of this paper  
is to generalize the formalism 
of the first part to recapture information about this noise that is lost in the 
passage to the system reduced density matrix.  We do this in Sec. 6 
by defining 
a density tensor hierarchy as the trace over the environment of a product 
of environment operators  constructed as the system matrix elements  of 
the total density matrix.  Unlike the classical noise construction, which 
uses only the system density matrix, the construction in the quantum noise 
case requires knowledge of the full system plus environment density matrix, 
and so (except for the order one case) does not give a system observable.  
It is nonetheless computable in any theory of the system plus environment, 
and is of theoretical, rather than empirical, interest.  Because the 
environment operators entering the construction are 
non-commutative, this hierarchy is no longer totally symmetric in its 
system index pairs, but by the cyclic permutation property of the 
trace, it is symmetric under cyclic permutation of the system index pairs. 
Also, because the system trace of these environment 
operators gives only the reduced environment density matrix, rather than 
unity, there is in general no descent equation associated with taking 
this trace.  However, when indices of adjacent system operators are 
contracted, one gets the square of the overall density matrix, and so there 
remains  a set of descent relations  connecting the order $(n)$ tensor to the 
order $(n-1)$ tensor.  Finally, in the case of thermal (or other mixed) 
overall states, we define the appropriate tensor  as a weighted 
sum of pure state tensors, in analogy with the definition of Sec. 2.  

In subsequent sections, we give applications of the trace 
hierarchy formalism to several classic problems discussed in the theory 
of quantum master equations.  In Sec. 7 we consider the quantum 
Brownian motion (and resulting decoherence) of a massive Brownian particle 
in interaction with an independent particle bath of scatterers.  In Sec. 8  
we discuss the tensor hierarchy corresponding to the weak coupling 
Born--Markov master equation, and it specialization to the quantum optical 
master equation.  Finally in Sec. 9, we give an analogous discussion for 
the Caldeira--Leggett model of a particle in interaction with a system 
of environmental oscillators.

We conclude with a discussion that bridges the considerations of the 
classical noise and the quantum noise cases.  In Sec. 10, we contrast 
two different It\^o stochastic Schr\"odinger equations, both of which 
have the same Lindblad, but only one of which leads to state vector 
reduction.  We relate this to the fact that the equation giving the  
time derivative of the stochastic  
expectation  of operator variances involves the order two density tensor,   
which differs for the two cases.  We discuss the analogous equation for 
the time dependence of the variance of a ``pointer operator'' in the 
case of a quantum system coupled to a quantum environment, and show why 
this does not lead to state vector reduction.  Thus we see no mechanism 
for quantum ``noise'' in a closed quantum system plus environment to 
provide a resolution of the quantum measurement problem.  
\bigskip
\centerline{{\bf 2.~~The density tensor for classical noise and its 
kinematical properties}}
\bigskip
We proceed to establish our notation and to define the density  tensor 
hierarchy in the classical noise case. 
We denote the pure state density matrix formed from 
the unit normalized state $|\psi_{\alpha}\rangle $ by $ \rho_{\alpha}$, with 
$$ \rho_{\alpha}=|\psi_{\alpha} \rangle \langle \psi_{\alpha} |~~~,
\eqno(3a)$$ 
and its general matrix element between states $|i\rangle $ and 
$|j \rangle $ of ${\cal H}_{\cal S}$ by 
$$ \rho_{\alpha;ij} \equiv \langle i|  \rho_{\alpha}|j \rangle~~~.\eqno(3b)$$
The unit trace condition on $ \rho_{\alpha}$ states that 
$${\rm Tr} \rho_{\alpha} = \langle \psi_{\alpha}|\psi_{\alpha} \rangle 
=1~~~,\eqno(3c)$$ 
and the idempotence condition on $ \rho_{\alpha}$ states that 
$$ \rho_{\alpha}^2= |\psi_{\alpha} \rangle \langle \psi_{\alpha} |
|\psi_{\alpha} \rangle \langle \psi_{\alpha} | =
|\psi_{\alpha} \rangle \langle \psi_{\alpha} |  = \rho_{\alpha}
~~~.\eqno(3d)$$
We now define the order $n$ density tensor by 
$$ \rho^{(n)}_{i_1j_1,i_2j_2,...,i_nj_n} 
= \sum_{\alpha} w_{\alpha}  \rho_{\alpha;i_1j_1} 
 \rho_{\alpha;i_2j_2}   
... \rho_{\alpha;i_nj_n}
=E[ \rho_{\alpha;i_1j_1} \rho_{\alpha;i_2j_2} 
... \rho_{\alpha;i_nj_n}]  
~~~,\eqno(4a)$$
with $E[F_{\alpha}]$ a shorthand for 
$$E[F_{\alpha}]=\sum_{\alpha} w_{\alpha} F_{\alpha}~~~.\eqno(4b)$$
Since 
$$\rho^{(1)}_{ij}=\sum_{\alpha} w_{\alpha}  \rho_{\alpha;ij} 
=\sum_{\alpha} w_{\alpha} \langle i| \rho_{\alpha} |j \rangle 
~~~,
\eqno(5a)$$ 
we see that this is just the $|i\rangle $ to 
$|j\rangle$ matrix element of the reduced density matrix $\rho$ defined 
in Eq.~(1b),
$$\rho^{(1)}_{ij}= \langle i|\rho|j \rangle~~~,\eqno(5b)$$ 
and so the density tensor of Eq.~(4a) is a natural generalization of the 
usual reduced density matrix.  When the states $|\psi_{\alpha}\rangle$ are 
independent of the label $\alpha$, the definition of Eq.~(4a) simplifies to 
$$\rho^{(n)}_{i_1j_1,i_2j_2,...,i_nj_n} =
\rho_{i_1j_1}\rho_{i_2j_2}...\rho_{i_nj_n}~~~,\eqno(5c)$$
and so the difference between Eq.~(4a) and a product of 
reduced density matrix 
elements is a reflection of the statistical structure of the ensemble.  
Since the factors within the expectation $E[...]$  on the right of Eq.~(4a) are just ordinary complex 
numbers, the density tensor is symmetric under interchange of any 
index pair $i_lj_l$ with any other index pair $i_mj_m$.  Consequently, 
we can define a generating function  
for the density tensor by 
$$G[a_{ij}]=E[e^{ \rho_{\alpha;ij} a_{ij}}]
=\sum_{n=0}^{\infty}{ a_{i_1j_1}...a_{i_nj_n} \over n!} 
\rho^{(n)}_{i_1j_1,...,i_nj_n}~~~,\eqno(5d)$$
where repeated indices $i,j$ are summed. It will often be convenient to  
abbreviate $\rho_{\alpha;ij} a_{ij}$ by $\rho_{\alpha} \cdot a$, so that 
the generating function becomes in this notation 
$G[a]=E[e^{\rho_{\alpha}\cdot a}]$.  

Although the 
density tensor for $n>1$ is not an operator on ${\cal H}_{\cal S}$, it 
clearly has the structure of an operator on the $n$-fold tensor product 
${\cal H}_{\cal S} \otimes {\cal H}_{\cal S} \otimes ...
\otimes {\cal H}_{\cal S}$. Motivated by this, we will often find it  
convenient to write the definition of Eq.~(4a) as 
$$\rho^{(n)}=E\left[ \prod_{\ell=1}^{n} 
\rho_{\alpha;\ell}\right]~~~,\eqno(5e)$$
with each factor $\rho_{\alpha;\ell}$  acting on a distinct factor 
Hilbert space ${\cal H}_{{\cal S}; \ell }$ in the tensor product 
$\prod_{\ell=1}^{n} {\cal H}_{{\cal S}; \ell }$.  One can pass easily 
back and forth from this notation to one in which the system matrix 
indices are displayed explicitly.  

Let us consider next the result of contracting any row index $i_l$ with 
any column index $j_k$.  There are two basic cases: (i) one can contract 
a row index $i_l$ with its corresponding column index $j_l$, and (ii) 
one can contract a row index $i_l$ with a column index $j_k$ with 
$k \not= l$.  Since the density tensor is symmetric in its index pairs, 
it suffices to consider only one example of each case, since all others  
can be obtained by permutation.  For the contraction of $i_1$ with $j_1$ 
we find  
$$\delta_{i_1j_1} \rho^{(n)}_{i_1j_1,i_2j_2,...,i_nj_n} 
= E[  ({\rm Tr} \rho)  \rho_{\alpha;i_2j_2}
... \rho_{\alpha;i_nj_n}]
=E[  \rho_{\alpha;i_2j_2} 
... \rho_{\alpha;i_nj_n} ]
=\rho^{(n-1)}_{i_2j_2,...,i_nj_n}
~~~,\eqno(6a)$$
where we have used the unit trace condition of Eq.~(3c).  For the 
contraction of $j_1$ with $i_2$, we find 
$$\delta_{j_1i_2} \rho^{(n)}_{i_1j_1,i_2j_2,...,i_nj_n} 
= E[   ( \rho^2)_{\alpha;i_1j_2}
... \rho_{\alpha;i_nj_n} ]
=E[  \rho_{\alpha;i_1j_2} \rho_{\alpha; i_3j_3} 
... \rho_{\alpha;i_nj_n} ]
=\rho^{(n-1)}_{i_1j_2,i_3j_3,...,i_nj_n}
~~~,\eqno(6b)$$
where now we have used the idempotence condition of Eq.~(3d).  As an 
illustration of how this works when all possible index pair contractions 
are considered, we give the complete set of contractions reducing 
the second order density tensor to a first order density tensor, 
$$\eqalign{
\delta_{i_1j_1} \rho^{(2)}_{i_1j_1,i_2j_2}=&\rho^{(1)}_{i_2j_2}~~~,\cr  
\delta_{i_2j_2} \rho^{(2)}_{i_1j_1,i_2j_2}=&\rho^{(1)}_{i_1j_1}~~~,\cr  
\delta_{j_1i_2} \rho^{(2)}_{i_1j_1,i_2j_2}=&\rho^{(1)}_{i_1j_2}~~~,\cr  
\delta_{j_2i_1} \rho^{(2)}_{i_1j_1,i_2j_2}=&\rho^{(1)}_{i_2j_1}~~~.\cr  
}\eqno(7a)$$
Referring to the generating function of Eq.~(5d), the general 
descent equations 
can be summarized compactly by the two identities, 
$$\eqalign{
\delta_{mr} {\partial G[a_{ij}] \over \partial a_{mr}}
=&E[({\rm Tr} \rho_{\alpha}) e^{ \rho_{\alpha;ij} a_{ij}}]   
=G[a_{ij}]~~~,\cr 
\delta_{rp}{\partial^2 G[a_{ij}]\over \partial a_{mr} \partial a_{pq} }
=& E[ \rho_{mr} \rho_{rq}  e^{ \rho_{\alpha;ij} a_{ij}}]   
=E[ \rho_{mq}  e^{ \rho_{\alpha;ij} a_{ij}}]   
={\partial G[a_{ij}] \over \partial a_{mq}} ~~~.\cr
}\eqno(7b)$$ 
When the density matrix $\rho$ used to define the density tensor is 
a mixed state density matrix, the trace descent relation of Eq.~(6a) is 
unchanged, while the indempotency relation of Eq.~(6b) relates the 
contraction an order $(n)$ tensor to an order $(n-1)$ tensor in which 
one factor $\rho$ is replaced by $\rho^2$; this is not a member of the 
original hierarchy, but still gives a useful relation for checking 
calculations.  

To conclude this section, let us return to the variances introduced 
by Breuer and Petruccione.  In terms of the order one and order 
two density tensors, we evidently have 
$$\eqalign{
{\rm Var}_1(R)=&\rho^{(1)}_{i_1j_1} (R^2)_{j_1i_1} 
-\rho^{(2)}_{i_1j_1,i_2j_2} R_{j_1i_1}R_{j_2i_2}~~~,\cr
{\rm Var}_2(R)=&\rho^{(2)}_{i_1j_1,i_2j_2} R_{j_1i_1}R_{j_2i_2} 
-(\rho^{(1)}_{i_1j_1} R_{j_1i_1})^2~~~,\cr     
{\rm Var}(R)=&\rho^{(1)}_{i_1j_1} (R^2)_{j_1i_1} 
-(\rho^{(1)}_{i_1j_1} R_{j_1i_1})^2~~~,\cr     
}\eqno(8a)$$
with $R_{ji} = \langle j|R|i \rangle$.  
Clearly, other statistical properties of the ensemble are readily expressed 
in terms of the density tensor hierarchy.  For example, the 
ensemble average of the product of the 
expectations of two different operators $R$ and $S$ is  
given by 
$$\sum_{\alpha} w_{\alpha} \langle \psi_{\alpha}|R|\psi_{\alpha} \rangle 
\langle \psi_{\alpha}|S|\psi_{\alpha} \rangle
=\rho^{(2)}_{i_1j_1,i_2j_2} R_{j_1i_1} S_{j_2i_2}~~~,\eqno(8b)$$
which can be used, together with information obtained from $\rho^{(1)}$,  
to calculate the covariance and correlation of $R$ and $S$.
\vfill\eject

\bigskip   
\centerline{\bf 3.~~Isotropic spin-1/2 ensemble}
\bigskip
As a simple example of the density tensor formalism, let us follow 
Breuer and Petruccione [1] and consider the case of an isotropic 
spin-1/2 ensemble.  Let $\vec v$ be a vector in three dimensions, 
and consider the ensemble of spin-1/2 pure state density matrices 
$$ \rho(\vec v)={1\over 2} (1+\vec v\cdot \vec \sigma)~~~,\eqno(9a)$$ 
with $\vec \sigma=(\sigma^1,\sigma^2,\sigma^3)$ the standard 
Pauli matrices, and with a uniform 
probability distribution of $\vec v$ over the unit sphere $|\vec v\,|=1$ 
specified by 
$$w(\vec v\,)={1\over 4 \pi} \delta(|\vec v\,|-1)~~~.\eqno(9b)$$
(Clearly, $\vec v$ has the same significance as the label $\alpha$ used 
in the preceding section.)  Defining 
$$E[P(\vec v\,)]  =\int d^3v w(\vec v\,) P(\vec v\,)~~~,\eqno(10a)$$ 
a standard calculation gives 
$$E[ 1] =1~,~~E[ v_s v_t ] ={1\over 3}\delta_{st}~,~...~~~,
\eqno(10b)$$
with all averages of odd powers of $\vec v$ vanishing.  From 
Eq.~(9a), we have 
$$ \rho(\vec v\,)_{ij}={1\over 2} (\delta_{ij}+v_r \sigma^r_{ij})~~~,
\eqno(11a)$$ 
and the general density tensor over this ensemble is defined by 
$$\rho^{(n)}_{i_1j_1,...,i_nj_n}=E[  \rho(\vec v\,)_{i_1j_1}
...\rho(\vec v\,)_{i_nj_n}]~~~.\eqno(11b)$$
From Eq.~(10b), 
the first three tensors in this hierarchy are now easily found to be 
$$\eqalign{
\rho^{(1)}_{i_1j_1}=&{1\over 2} \delta_{i_1j_1}~~~,\cr
\rho^{(2)}_{i_1j_1,i_2j_2}=&{1\over 4}\left(\delta_{i_1j_1}\delta_{i_2j_2} 
+{1\over 3} \vec \sigma_{i_1j_1}\cdot \vec \sigma_{i_2j_2}\right)~~~,\cr
\rho^{(3)}_{i_1j_1,i_2j_2,i_3j_3}=&{1\over 8} \left[
\delta_{i_1j_1}\delta_{i_2j_2}\delta_{i_3j_3}
+{1\over 3} \left(\delta_{i_1j_1}\vec \sigma_{i_2j_2} \cdot \vec \sigma_{i_3 j_3} 
+\delta_{i_2j_2} \vec \sigma_{i_1j_1}\cdot \vec \sigma_{i_3j_3} 
+\delta_{i_3j_3} \vec \sigma_{i_1j_1} \cdot \vec \sigma_{i_2j_2}\right) 
\right]~~~.\cr  
}\eqno(12)$$
Using the relations ${\rm Tr} \vec \sigma=0$ and $(\vec \sigma^{\,2})_{ij}  
=3 \delta_{ij}$, it is now easy to verify that the descent relations of 
Eqs.~(6a) and (6b) are satisfied by Eq.~(12).

For the isotropic spin-1/2 ensemble, the generating function of Eq.~(5d) 
becomes 
$$G[a_{ij}]=E[e^{ \rho(\vec v)_{ij}a_{ij}}]~~~, \eqno(13)$$
with $ \rho(\vec v\,)_{ij}$ given by Eq.~(11a).  
Defining the vector $\vec A$ by 
$$\vec A={1\over 2}\vec \sigma_{ij} a_{ij}~~~,\eqno(14a)$$
a simple calculation gives 
$$G[a_{ij}]=\exp({1\over 2} {\rm Tr}a)  {\sinh |\vec A\,| \over |\vec A\,|}
= \exp({1\over 2} {\rm Tr}a) [1 + {\vec A^{\,2} \over 3!} 
+ {(\vec A^{\,2})^2 \over 5!}+...]~~~,\eqno(14b)$$
from which one can read off the values of the low order density tensors 
given in Eq.~(12).  The verification of the descent relations of Eq.~(7b) 
for the generating function of Eq.~(14b) is given in Appendix A.  
\bigskip

\centerline{\bf 4.~~It\^o stochastic Schr\"odinger equation} 
\bigskip
We consider next a state vector $|\psi\rangle$ with a 
time evolution described by a stochastic Schr\"odinger equation, which is   
a frequently used model approximation to open system dynamics.  In this 
case the state vector and the corresponding pure state density matrix 
$ \rho=|\psi\rangle \langle \psi|$ are implicit functions of the 
noise, which takes a different sequence of values for each history of the 
system.  In the notation of Sec. 2, the different histories are labeled by 
the subscript $\alpha$, and the expectation of Eq.~(4b) is an average  
over all possible histories.  It is customary, however, in discussing 
stochastic Schr\"odinger equations to omit the subscript $\alpha$, 
treating the history dependence of $ \rho$ as understood.  So 
in this context, the definition of Eq.~(4a) becomes 
$$\rho^{(n)}_{i_1j_1,...,i_nj_n}=E[ \rho_{i_1j_1}... \rho_{i_nj_n}]
~~~,\eqno(15)$$ 
with $E[...]$ the stochastic expectation, and the generating function 
$G[a_{ij}]$ takes the same form as given in Eq.~(5d) but with the 
subscript $\alpha$ omitted.    

Our aim in this section is to derive an equation of motion for the generating 
function, which on expansion yields equations of motion for all density  
tensors $\rho^{(n)}$, taking as input the general pure state density 
matrix evolution constructed by Wiseman and Di\'osi [7], 
that corresponds to a given 
Lindblad form  [8,9] for the time evolution of the reduced density matrix 
$\rho^{(1)} =E[\rho]$.  We begin by recapitulating the results of ref [7].  
The most general evolution of a density matrix $\rho$ that 
preserves ${\rm Tr}\rho=1$ and obeys the complete positivity condition 
is the Lindblad form 
$$d \rho = dt {\cal L} \rho ~~~,\eqno(16a)$$
with 
$${\cal L} \rho \equiv -i[H_{\rm TOT},\rho] + c_k\rho c_k^{\dagger} -{1\over 2} 
\{c_k^{\dagger}c_k,\rho\}~~~,\eqno(16b)$$ 
with $\{,\}$ denoting the anticommutator, and with the repeated index 
$k$ summed.  The set of Lindblad operators $c_k$ describes the effects 
on the system 
of the reservoir or environment that is modeled by an external 
classical noise. 
Wiseman and Di\'osi show that the most general evolution of 
the pure state density matrix $ \rho$ for which $E[d\rho]$ 
reduces to Eqs. (16a) and (16b) takes  the form 
$$d \rho=dt {\cal L} \rho+|d\phi\rangle\langle\psi| 
+ |\psi \rangle \langle d\phi|~~~.\eqno(17a)$$
Here $|d\phi\rangle$ is a state vector that is a pure noise term, so that 
$$E[|d\phi\rangle]=0~~~,\eqno(17b)$$ 
that is orthogonal to $|\psi\rangle$, so that 
$$\langle \psi|d\phi\rangle =0~~~,\eqno(17c)$$ 
and that obeys 
$$|d\phi\rangle \langle d\phi|=dt W~~~. \eqno(17d)$$ 
The operator  $W$ is the Di\'osi transition rate operator [5] given by 
$$\eqalign{
W=&{\cal L}\rho-\{\rho,{\cal L}\rho\}+ \rho {\rm Tr}( \rho {\cal L} \rho)\cr
=&(c_k-\langle c_k\rangle) \rho (c_k-\langle c_k\rangle)^{\dagger} \cr
}~~~,\eqno(18)$$
where $\langle c_k\rangle $ is a shorthand for 
the quantum state expectation $\langle \psi|c_k|\psi \rangle 
={\rm Tr} \rho c_k$.  Although $|d\phi \rangle \langle d\phi|$ is 
completely fixed, Wiseman and Di\'osi show that $|d\phi\rangle |d\phi \rangle$ 
is free, with different choices for this and different phase choices for 
the $c_k$ corresponding to different  
pure state evolutions (or ``unravelings'') that yield the same evolution 
of Eqs.~(16a) and (16b) for the reduced density matrix $\rho$.  

Wiseman and Di\'osi further show that $|d\phi\rangle$ can be parameterized 
by complex Wiener processes by writing 
$$|d\phi\rangle=(c_k-\langle c_k \rangle) |\psi\rangle d\xi_k^*
~~~,\eqno(19a)$$
with 
$$E[d\xi_k]=E[d\xi_k^*]=0   ~~~\eqno(19b)$$ 
and with
$$\eqalign{
d\xi_j(t)d\xi_k^*(t)=&dt \delta_{jk}  \cr
d\xi_j(t)d\xi_k(t)=&dt u_{jk}~~~,\cr
}\eqno(19c)$$
where $u_{kj}=u_{jk}$ is a set of arbitrary complex numbers subject to the 
condition that the norm of the complex matrix ${\bf u}\equiv [u_{jk}]$
be less than or equal to 1.  (See Eqs.~(4.10) and (4.11) of ref. [7].) In 
terms of this parameterization of $|d\phi\rangle$, the pure state evolution 
of Eq.~(17a) takes the form 
$$d \rho=dt {\cal L}\rho  + (c_k-\langle c_k\rangle) \rho d\xi_k^*
+ \rho (c_k-\langle c_k\rangle)^{\dagger} d\xi_k~~~,\eqno(19d)$$
and the corresponding stochastic Schr\"odinger equation for the wave 
function is [7] 
$$\eqalign{ 
d|\psi\rangle =&-iH_{\psi}dt |\psi\rangle + (c_k-\langle c_k\rangle)d\xi_k^* 
|\psi \rangle~~~, \cr
-iH_{\psi}=& -iH_{\rm TOT}-{1\over 2}\big(c_k^{\dagger}c_k-2\langle c_k \rangle^* c_k 
+\langle c_k \rangle^* \langle c_k \rangle  \big)~~~.\cr
}\eqno(19e)$$

We proceed now to use pure state evolution of Eq.~(19d) to calculate the 
evolution equation for the generating function 
$$G[a_{ij}]=E[\exp(\rho_{ij} a_{ij})]~~~.\eqno(20a)$$   
To calculate the differential of Eq.~(20a), we use the It\^o stochastic 
calculus rule for the differential of a function $f(w)$ of a stochastic 
variable $w$, 
$$df(w)=dwf^{\prime}(w) + {1\over 2} (dw)^2 f^{\prime\prime}(w)~~~.\eqno(20b)$$
Applying this to Eq.~(20a), we get 
$$dG[a_{ij}]=E[(d\rho_{mr}a_{mr} + {1\over 2} 
d\rho_{mr}a_{mr}d\rho_{pq}a_{pq})
\exp(\rho_{ij} a_{ij})]~~~.\eqno(20c)$$   
Substituting Eq.~(19d) for $d \rho$, and using Eqs.~(19a-c), together 
with the It\^o calculus rule $E[dwf(w)]=0$, we get 
$$dG[a_{ij}]=dt E\big[\big(a_{mr}({\cal L}\rho)_{mr}           
+{1\over 2}a_{mr}a_{pq} C_{mr,pq}\big) \exp(\rho_{ij} a_{ij})\big]~~~,
\eqno(21a)$$   
with the coefficient of the quadratic term in $a_{ij}$ given by 
$$\eqalign{ 
C_{mr,pq} =&C_{pq,mr}=d\rho_{mr}d\rho_{pq} \cr
=&\langle m| (c_k-\langle c_k \rangle)  \rho |r\rangle
 \langle p| \rho(c_k-\langle c_k\rangle)^{\dagger}|q\rangle  \cr
+&\langle m| \rho (c_k-\langle c_k\rangle)^{\dagger} |r\rangle
 \langle p|  (c_k-\langle c_k \rangle)  \rho |q\rangle \cr
+&\langle m|  (c_k-\langle c_k \rangle)  \rho |r\rangle
 \langle p|(c_{\ell}-\langle c_{\ell}\rangle) \rho|q\rangle 
 u^*_{k\ell} \cr
+&\langle m|  \rho(c_k-\langle c_k\rangle)^{\dagger}|r\rangle 
 \langle p|   \rho(c_{\ell}-\langle c_{\ell}\rangle)^{\dagger}|q\rangle 
 u_{k \ell} \cr
}\eqno(21b)$$
This expression can be rearranged by using the identity, valid for general 
operators $A,B$, general states $|r\rangle,|m\rangle$,  and general 
pure state (idempotent) density matrix $\rho$, 
$$ \rho A |r\rangle \langle m|B \rho=
 \rho \langle m|B  \rho A|r\rangle ~~~, \eqno(22a)$$
giving an alternative result for $C_{mr,pq}$ 
$$\eqalign{
C_{mr,pq}=&W_{mq}  \rho_{pr}+ W_{pr} \rho_{mq} \cr 
+&[(c_k-\langle c_k\rangle) \rho]_{mq}u^*_{k\ell}
[(c_{\ell}-\langle c_{\ell} \rangle) \rho]_{pr} \cr  
+&[ \rho (c_k-\langle c_k \rangle)^{\dagger}]_{pr} u_{k\ell}
[ \rho (c_{\ell}-\langle c_{\ell} \rangle)^{\dagger}]_{mq}~~~, \cr
}\eqno(22b)$$
where we have used Eq.~(18) defining the operator $W$, and where we use 
the subscript notation of Eq.~(3b) for matrix elements, so that in general 
$A_{mr}=\langle m|A|r \rangle $.  

From the evolution equation of Eqs. (21a,b) and (22b) for 
the generating function, by expansion in powers of $a$ we can read off 
the evolution equation for the general density tensor of order $n$.  
Employing now the condensed notation of Eq.~(5e), in which matrix indices 
are not indicated explicitly, we have 
$$\eqalign{
d\rho^{(n)}=&dt E[\sum_{\ell=1}^n (\rho_1...\rho_n)_{\ell}  
({\cal L}\rho)_{\ell} \cr
+& \sum_{\ell < m=1}^n (\rho_1...\rho_n)_{\ell m} C_{\ell m}]~~~.\cr 
}\eqno(23a)$$
Here $(\rho_1...\rho_n)_{\ell}$ denotes the product 
$\prod_{j=1}^n  \rho_j$ with the factor $\rho_{\ell}$ omitted, and similarly, 
$(\rho_1...\rho_n)_{\ell m}$ denotes the product $\prod_{j=1}^n  \rho_j$
with the factors $\rho_{\ell}$ and $\rho_m$ omitted.\footnote{$^2$}  
{For $n=1$, $(\rho_1)_1=1$ and $(\rho_1)_{\ell m}=0$, while for $n=2$, 
$(\rho_1\rho_2)_{12}=1$.}
The coefficient 
$C_{\ell m}$ is given by 
$$\eqalign{ 
C_{\ell m}  =C_{m\ell}=&[ (c_k-\langle c_k \rangle)  \rho ]_{\ell}
 [ \rho(c_k-\langle c_k\rangle)^{\dagger}]_m  \cr
+&[ \rho (c_k-\langle c_k\rangle)^{\dagger}]_{\ell} 
 [ (c_k-\langle c_k \rangle)  \rho ]_m \cr
+&[  (c_k-\langle c_k \rangle)  \rho ]_{\ell}
[(c_{\bar k}-\langle c_{\bar k}\rangle) \rho]_m 
 u^*_{k\bar k} \cr
+&[  \rho(c_k-\langle c_k\rangle)^{\dagger}]_{\ell}  
 [ \rho(c_{\bar k}-\langle c_{\bar k}\rangle)^{\dagger}]_m 
 u_{k \bar k}~~~, \cr
}\eqno(23b)$$
which corresponds in an obvious way to Eq.~(21b) when matrix elements 
are written explicitly between states $\langle m| $ and $|r \rangle$ in   
the Hilbert space labeled by $\ell$, and between states $\langle p|$ and $|q \rangle$ 
in the Hilbert space labeled by $m$.  (No relation is implied between 
the $m$ used as a state label, and the $m$ used as a Hilbert space label.)
Since  $C_{\ell m}$ in  Eq.~(23a), 
which depends through the terms involving $u_{k\bar k}$ on the choice 
of unraveling, 
is multiplied by two powers of $a$, it does not 
contribute to the evolution equation for the reduced density matrix 
$\rho^{(1)}$.  So as expected, the reduced density matrix evolution is 
given solely by the Lindblad term and is independent of the choice of 
unraveling.  Higher density tensors $\rho^{(n)}$, with $n \geq 2$, 
have evolution equations that receive contributions from $C_{\ell m}$, 
and so contain information that distinguishes between different 
unravelings of the Lindblad evolution.  

As a simple illustration of how the tensors $\rho^{(n)}$ for $n\geq 2$ 
distinguish between different unravelings, let us consider the case of 
real noise,  $d\xi_k=d\xi_k^*$, for which $u_{jk}=\delta_{jk}$, and with   
a single Lindblad $c_1$, which we choose as either $c_1=A$ or $c_1=iA$, 
with $A$ a self-adjoint operator.  Both choices of $c_1$ lead to the 
same Lindblad, since ${\cal L}$ is invariant under rephasing of  
$c_k$, but through the $u_{k \bar k}$ terms they lead to different  
expressions for $C_{mr,pq}$.  When $c_k=A$, we find from Eq.~(21b) 
$$\eqalign{
C_{mr,pq}
=&\langle m| \{A-\langle A \rangle, \rho\} |r\rangle
\langle p|  \{A-\langle A \rangle, \rho\} |q\rangle   \cr
=&\langle m| [\rho,[\rho,A]] |r\rangle
\langle p| [\rho,[\rho,A]]  |q\rangle   
~~~,\cr
}\eqno(24a)$$
while when $c_k=iA$, we have instead 
$$C_{mr,pq}=-\langle m| [A, \rho] |r\rangle
\langle p| [A, \rho] |q\rangle~~~.\eqno(24b)$$
We will return to this example in Sec. 10.  

Using the expression of Eq.~(21a) for the time evolution of the 
generating function, the 
descent equations of Eq.~(7b) can be verified; this calculation is 
carried out in Appendix B. 
\bigskip

\centerline{\bf 5. Jump process Schr\"odinger equation} 
\bigskip
As our next density tensor application we consider the jump process 
(piecewise deterministic process, or PDP) Schr\"odinger equation, given 
by 
$$d|\psi \rangle= Adt |\psi \rangle + B_k dN_k |\psi\rangle~~~,\eqno(25a)$$
where a sum over $k$ is understood, 
with $A$ and the $B_k$ general (non-self-adjoint) operators, and with the 
$dN_k$ independent discrete random variables obeying 
$$dN_jdN_k=\delta_{jk} dN_k~,~~dN_j dt=0~~~. \eqno(25b)$$
Straightforward calculation shows that this process preserves the 
norm of $|\psi\rangle$ and the pure state condition 
$ \rho^2= \rho=|\psi \rangle \langle \psi|$, 
provided that $A$ and $B$ obey the restrictions 
$$\eqalign{ 
\langle A + A^{\dagger}\rangle =&0,~~~ \cr
\langle B_k + B_k^{\dagger}+B_k^{\dagger}B_k\rangle=&0 ~~~,\cr
}\eqno(25c)$$ 
with no summation over $k$ on the second line, which must hold 
individually for each value of $k$.  Corresponding to Eq.~(25a),   
the density matrix obeys the evolution equation 
$$\eqalign{ 
d\rho=&(A\rho+\rho A^{\dagger}) dt + Q_k dN_k~~~,\cr
Q_k=&B_k \rho +\rho B_k^{\dagger} 
+B_k \rho B_k^{\dagger}~~~,\cr 
}\eqno(25d)$$ 
with a sum over $k$ understood in the $dN_k$ term on the first line,     
but no sum over $k$ understood in the second line.  

Let now $E_{|\psi\rangle}[...]$ denote an expectation conditioned on the current 
value of the wave function being $|\psi\rangle$, and $E[...]$ be   
the expectation value over the entire history of the jump process (which 
leads to an ensemble of different current values of the wave function). 
We wish to find 
restrictions on $A$, $B_k$, and on 
$$E_{|\psi\rangle}[dN_k]\equiv v_k dt~~~,\eqno(26a)$$
such that the expectation of $d\rho$ takes the Lindblad form of Eq.~(16b), 
that is, 
$$\eqalign{
E[d\rho]=&dt {\cal L} \rho\cr
{\cal L}\rho=&
 -i[H_{\rm TOT},\rho] + c_k\rho c_k^{\dagger} -{1\over 2} 
\{c_k^{\dagger}c_k,\rho\}~~~.\cr
}\eqno(26b)$$ 
Making the Ansatz 
$$B_k={c_k -K_k \over v_k^{1\over 2} }-1~~~,\eqno(27a)$$
with $K_k$ constants (this Ansatz includes both the standard quantum jump 
equation ($K_k=0$), and the 
orthogonal jump equation ($K_k=\langle c_k \rangle$), as special cases; see 
 Schack and Brun [10] for a concise review), some calculation shows that   
the conditions of Eqs.~(25c) and (26b) are satisfied if we choose  
$$\eqalign{
v_k=&\langle (c_k-K_k)^{\dagger}(c_k-K_k) \rangle ~~~,\cr
A=&-iH_{\rm TOT}-{1\over 2} c_k^{\dagger}c_k 
+ {1\over 2} \langle c_k^\dagger c_k \rangle
+c_k K_k^* -{1\over 2} (\langle c_k\rangle K_k^* +\langle c_k\rangle^* K_k )
~~~.\cr
}\eqno(27b)$$

Let us now define the order $n$ density tensor for the jump models by 
$$\rho^{(n)}=E[\prod_{\ell=1}^n \rho_{\ell}]~~~,\eqno(28a)$$ 
where we use the condensed notation of Eq.~(5e).  For the differential  
of this, we find
$$\eqalign{
d\rho^{(n)}=&E[\sum_{\ell=1}^n (\rho_1...\rho_n)_{\ell} d\rho_{\ell} 
+\sum_{\ell<m=1}^n (\rho_1...\rho_n)_{\ell m} d\rho_{\ell} d\rho_m  \cr
+& \sum_{\ell<m<p=1}^n (\rho_1...\rho_n)_{\ell m p}  
d\rho_{\ell} d\rho_m d\rho_p 
+...+d\rho_1d\rho_2d\rho_3...d\rho_{n-1} d\rho_n]~~~~,\cr 
}\eqno(28b)$$ 
where all powers of $d\rho$ must be retained because $dN_k^2=dN_k$. 
Using the conditional probability formula 
$p(|\psi\rangle \cap dN_k)=p(dN_k|~|\psi\rangle)  
p(|\psi\rangle)$, we get the conditional expectation formula, valid 
for an arbitrary 
function $F$ of the state $|\psi\rangle$, 
$$E[F(|\psi\rangle) dN_k]=E[F(|\psi\rangle) E_{|\psi\rangle} [dN_k] ]
=E[F(|\psi\rangle)v_k]~~~.\eqno(29a)$$ 
Using this equation to evaluate the higher order terms in Eq.~(28b), together 
with Eq.~(26b) for the leading term, we get 
$$\eqalign{
d\rho^{(n)}=&dt E[\sum_{\ell=1}^n (\rho_1...\rho_n)_{\ell} 
({\cal L}\rho)_{\ell} \cr 
+&\sum_{\ell<m=1}^n (\rho_1...\rho_n)_{\ell m} v_k (Q_k)_{\ell} (Q_k)_m \cr
+& \sum_{\ell<m<p=1}^n (\rho_1...\rho_n)_{\ell m p} 
v_k (Q_k)_{\ell} (Q_k)_m (Q_k)_p
+...+v_k (Q_k)_1 (Q_k)_2 (Q_k)_3 ...(Q_k)_{n-1}(Q_k)_n]       ~~~~,\cr 
}\eqno(29b)$$ 
with a sum over $k$ in each term containing $v_k$.  

Writing the corresponding generating function in compact notation as 
$$G[a]=E[e^{a\cdot \rho}]~~~,\eqno(30a)$$
the evolution equation for $G$ is given , with the $k$ sum now 
indicated explicitly, by 
$$\eqalign{
dG[a]=& E[e^{a \cdot d\rho} e^{a \cdot \rho}] - E[ e^{a \cdot \rho}] \cr
=&E[(\sum_{p=1}^{\infty}  {(a \cdot d\rho)^p \over p!})  e^{a \cdot \rho}] \cr
=&
dt E[(a \cdot {\cal L} \rho + \sum_{p=2}^{\infty} \sum_k v_k 
{(a \cdot Q_k)^p \over p!} ) e^{a \cdot \rho}] ~~~.\cr
}\eqno(30b)$$
From Eq.~(30b), and the identities \big( which follow, after some algebra, 
from Eqs.~(16b), (25d), (27a), and (27b) \big)
$$\eqalign{
\{\rho,{\cal L}\rho\} =&{\cal L}\rho -\sum_k v_k Q_k^2~~~,\cr
\{\rho,Q_k\}=&Q_k-Q_k^2~~~,\cr
}\eqno(30c)$$
one can prove that Eq.~(30b) obeys the descent equations, as 
shown in Appendix C.  
\bigskip

\centerline{\bf 6.~~The density tensor for quantum noise and its 
kinematical properties} 
\bigskip
Let us now consider a closed quantum system, consisting of 
a system ${\cal S}$ interacting with an environment ${\cal E}$. 
 In such 
a situation, one does not have a classical probability distribution 
$w_{ \alpha}$ and fluctuations associated 
with this probability distribution. 
Instead, one deals with the system plus environment as the only pure state  
that is given, with the fluctuations that are averaged over in deriving 
the master equation coming  from quantum fluctuations associated with 
the system-environment interaction.  Weighted averages of the sort 
that we have used in our definition of Eq.~(4a) appear only when the 
total state is a mixture of pure states, such as a thermal state, but 
in this case, 
important system quantum fluctuations  still occur in each pure 
state component of this mixture.  In order to describe this more 
general situation, we 
shall have to generalize our definition of a density tensor hierarchy.

To achieve this,   
we initially suppose the overall system plus environment to 
have the pure state density matrix   
$\rho$. We  denote 
the system basis states by $|i\rangle$ , as well as $|j \rangle$, 
and denote the environment basis states by $|e_a\rangle, a=1,2,....$. A general  
density matrix element has the form $\langle e_1 i| \rho |e_2 j\rangle$, and 
the standard reduced density matrix, with the environment traced out, 
is defined by 
$$\rho^{(1)}_{ij}=({\rm Tr}_{\cal E} \rho )_{ij}= 
\sum_e \langle e i|\rho|e j \rangle~~~.\eqno(31)$$ 
In order to recapture fluctuations that are averaged over in the trace 
in Eq.~(31), we define the density tensor $\rho^{(n)}$ by 
$$\eqalign{
\rho^{(n)}_{i_1j_1,i_2j_2,...,i_nj_n} =&
\sum_{e_1,e_2,...,e_n} \langle e_1 i_1| \rho |e_2 j_1\rangle
\langle e_2 i_2| \rho |e_3 j_2\rangle ...
\langle e_{n-1} i_{n-1}| \rho |e_n j_{n-1}\rangle
\langle e_n i_n| \rho |e_1 j_n\rangle  \cr
=&{\rm Tr}_{\cal E} \rho_{i_1j_1} \rho_{i_2j_2}...\rho_{i_nj_n}      ~~~. \cr
}\eqno(32a)$$
Here we have defined $\rho_{i_{\ell} j_{\ell}}$ as the matrix, labeled  
by the system state labels $i_{\ell},\,j_{\ell}$, acting 
on the environment Hilbert space ${\cal H}_{\cal E}$ according to  
$$ (\rho_{i_{\ell} j_{\ell} })_{e_1e_2} = 
\langle e_1|\rho_{i_{\ell} j_{\ell}}|e_2\rangle =
\langle e_1 i_{\ell}|\rho|e_2 j_{\ell} \rangle~~~.\eqno(32b)$$ 
The density tensor $\rho^{(n)}$ is again an operator on a tensor product 
of system Hilbert spaces $\prod_{\ell=1}^n  {\cal H}_{{\cal S};\ell}$.  
Thus, in a condensed notation analogous to that of Eq.~(5e), 
we can also write Eq.~(32a) as 
$$\rho^{(n)}={\rm Tr}_{\cal E} \rho_1 \rho_2....\rho_n~~~,\eqno(32c)$$
where $\rho_{\ell}$ is an operator acting on ${\cal H}_{\cal E} \otimes 
{\cal H}_{{\cal S};\ell}$.  

We have avoided using a product notation 
$\prod_{\ell=1}^n$ in Eq.~(32c) because the factors $\rho_{i_{\ell}j_{\ell}}$ in 
Eq.~(32a) and $\rho_{\ell}$ in Eq.~(32c) are different operators on the 
environment for each $\ell$ and thus do not commute.  Hence 
the density tensor is not symmetric under permutation of its pair indices 
$i_{\ell}j_{\ell}$, but it is symmetric under cyclical permutation of the 
indices, as a result of the cyclic symmetry of the trace.  For $n=2$, 
cyclic symmetry is equivalent to symmetry under pair index interchange, 
and for $n=3$, using the identity 
$${\rm Tr}ABC ={\rm Tr}{1\over 2}( [A,B]C 
+\{A,B\}C)~~~,\eqno(33)$$ 
cyclic symmetry is equivalent to the statement that the 
density tensor $\rho^{(3)}$ can be written as the sum of two 
tensors $\rho^{(3)}=\rho^{(3S)}+\rho^{(3A)}$, with $\rho^{(3S)}$  
completely symmetric, and $\rho^{(3A)}$ completely antisymmetric, under 
pair index interchange.   
Also because the density tensor is not totally symmetric in its pair indices, 
we cannot introduce a generating function by imitating Eq.~(5d)

Similarly, because of factor non-commutativity, the density tensor satisfies 
only a subset of the descent equations of Eqs.~(6a), (6b), and (7b).  
Contraction with $\delta_{i_{\ell}j_{\ell}}$ does not lead to a descent 
condition, since $\delta_{i_{\ell}j_{\ell}} \rho_{i_{\ell} j_{\ell}}$ is  
not unity, but rather ${\rm Tr}_{\rm S}\rho$, the reduced density matrix 
that acts on the environment when the system is traced out.  Contraction 
of a general  $j_k$  with a general $i_{\ell}$  for $k \not= \ell$ gives 
nothing useful, since in general non-commuting factors stand between 
$\rho_k$ and $\rho_{\ell}$.  However, when a column index $j_k$ is 
contracted with the  adjacent row index $i_{k+1}$, the two density 
matrices to which they are attached are linked to form the product 
$\rho^2=\rho$, and so we get the descent relation of Eq.~(6b), and 
others related to it by cyclic permutation symmetry, 
$$\delta_{j_1i_2}\rho^{(n)}_{i_1j_1,i_2j_2,...,i_nj_n}=
\rho^{(n-1)}_{i_1j_2,i_3j_3,...,i_nj_n}~~~.\eqno(34)$$
As noted before, even when $\rho^2 \not= \rho$, the descent relation 
corresponding to Eq.~(34) is still useful for checking calculations.
Since we cannot define a generating function as in Eq.~(5d), 
in the quantum noise 
case we do not have analogs of the descent equations in the form of Eq.~(7b); 
when verifying the descent equations in the various cases considered 
below, we will work directly from Eq.~(34).  

We will also consider a more general definition of the density tensor, 
corresponding to the case in which the system plus environment is in 
a mixed state composed of pure states $\rho_{\alpha}$ with weights 
$w_{\alpha}$.  Typically, $\alpha$ refers to an eigenvalue of a conserved 
quantum number of the total system, such as the energy; when the environment 
is considered in the independent particle approximation, with the system 
back reaction on the environment neglected, $\alpha$ then can 
refer to the energies and momenta of each environmental particle.   
In this case we define the density tensor by 
$$
\rho^{(n)}_{i_1j_1,i_2j_2,...,i_nj_n} =
\sum_{\alpha} w_{\alpha} \rho^{(n)}_{\alpha;i_1j_1,i_2j_2,...,i_nj_n} ~~~,
\eqno(35a)$$
with 
$$ \rho^{(n)}_{\alpha;i_1j_1,i_2j_2,...,i_nj_n}=
{\rm Tr}_{\cal E} \rho_{\alpha;i_1j_1} \rho_{\alpha;i_2j_2}...
\rho_{\alpha;i_nj_n} ~~~.\eqno(35b)$$
This definition gives information about both the quantum noise or 
fluctuations contained within each $\rho_{\alpha}$, and the classical 
noise or fluctuations associated with the probability distribution 
$w_{\alpha}$.   Note that in the mixed state case one could also 
define a density tensor that is a direct analog of the classical 
noise definition of Sec. 2, by 
$$
\rho^{(n);{\rm CL}}_{i_1j_1,i_2j_2,...,i_nj_n} =
\sum_{\alpha} w_{\alpha} \prod_{\ell=1}^n {\rm Tr}_{\cal E} 
\rho_{\alpha;i_{\ell}j_{\ell}} ~~~\eqno(36)$$
which would give information only about the classical noise fluctuations 
associated with the probability distribution $w_{\alpha}$.  In the examples  
computed in the following sections, where a weak coupling approximation is 
made,  the definition of Eq.~(36) typically contains no more information 
than could be gotten from a product  of $n$
reduced density matrix factors, each of the form 
$\sum_{\alpha} w_{\alpha}  {\rm Tr}_{\cal E} \rho_{\alpha;i_{\ell}j_{\ell}}$.

As already noted, the density tensor $\rho^{(n)}$ is not  measurable by any operation on the system Hilbert space.  Its construction requires 
knowledge of the full system plus environment density matrix, which 
is not experimentally  accessible for complex environments. Nonetheless $\rho^{(n)}$ is computable 
in any theory of the system-environment interaction, and we believe 
it to be of conceptual and theoretical interest, even if not of 
direct empirical relevance.

We close out this section by noting that in the quantum noise case, 
there is no analog of Eq.(8a), which relates the positive semidefinite 
variations ${\rm Var}_{1,2}$ to the density tensor $\rho^{(2)}$ in the classical 
noise case.   The
closest analog we  find to the fluctuation formulas of 
Eq.~(8a) involves the $n=3$ density tensor.   The reason for this 
is that whereas $E[1]=1$, the trace over the environment of unity is the 
dimension of the environmental Hilbert space; to get a unit trace over the 
environment we must include a factor of 
$ \rho_{\cal E}\equiv {\rm Tr}_{\cal S} 
 \rho$, the reduced density matrix for the environment.  This pushes up 
the order of the density tensor involved from 2 to 3.  Specifically, 
let $A_{\cal S}$ be an operator acting on ${\cal H}_{\cal S}$, but which 
acts as the unit operator on ${\cal H}_{\cal E}$.  In place of the   
expectations used in the classical noise discussion of 
Eqs.~(1a) through (2d),    
in the quantum noise case of system plus environment we consider   
the expression  $A_{\cal E} \equiv {\rm Tr}_{\cal S} 
\rho A_{\cal S}$, which is an operator on the environmental Hilbert  
space.  The trace of this operator over the environment is 
${\rm Tr}_{\cal E} A_{\cal E} 
={\rm Tr}_{\cal E}{\rm Tr}_{\cal S}  \rho  A_{\cal S}  
={\rm Tr}_{\cal S} ({\rm Tr}_{\cal E} \rho)  A_{\cal S}  
= {\rm Tr}_{\cal S} \rho^{(1)} A_{\cal S}$, giving the expectation of   
the operator $A_{\cal S}$ when the environment is not observed. 
On the other hand, the expectation of this operator formed from the  
environmental reduced density matrix 
is  ${\rm Tr}_{\cal E}\rho_{\cal E}  A_{\cal E}$.   
The mean squared fluctuation 
of this operator over the environment is positive semidefinite, 
and is given by  
$${\rm Tr}_{\cal E}  \rho_{\cal E} (A_{\cal E} - 
{\rm Tr}_{\cal E}  \rho_{\cal E} A_{\cal E})^2 
={\rm Tr}_{\cal E}  \rho_{\cal E} A_{\cal E}^2 
-({\rm Tr}_{\cal E}  \rho_{\cal E} A_{\cal E})^2~~~, \eqno(37a)$$
where we have used the fact that ${\rm Tr}_{\cal E} \rho_{\cal E} 
={\rm Tr}  \rho =1$.  Reexpressing Eq.~(37a) entirely in terms of 
the pure state density matrix $ \rho$, we have 
$$\eqalign{
&{\rm Tr}_{\cal E} {\rm Tr}_{\cal S}  \rho 
({\rm Tr}_{\cal S} \rho A_{\cal S})^2
-({\rm Tr}_{\cal E} {\rm Tr}_{\cal S}  \rho {\rm Tr}_{\cal S} 
 \rho A_{\cal S})^2 \cr
=&\delta_{j_1i_1} A_{{\cal S}j_2i_2}A_{{\cal S}j_3i_3} 
\rho^{(3S)}_{i_1j_1,i_2j_2,i_3j_3} -(\delta_{j_1i_1}A_{{\cal S}j_2i_2} 
\rho^{(2)}_{i_1j_1,i_2j_2})^2~~~,\cr
}\eqno(37b)$$
where we have used the fact that the right hand side of Eq.~(37b) involves 
only the symmetric part of the order 3 density tensor.  
Thus, as noted above, where 
a $n=2$ density tensor appears in Eq.~(8a), a $n=3$ 
density tensor appears in Eq.~(37b), and where 
a $n=1$ density tensor appears in Eq.~(8a), a $n=2$ 
density tensor appears in Eq.~(37b).     
\bigskip
\centerline{\bf 7.~~Collisional Brownian Motion}
\bigskip
As our first application of Eqs.~(32a-c) and Eqs.~(35a,b), we consider 
the collisional Brownian motion of a massive Brownian particle immersed 
in a bath of scattering particles.   We work in the approximation of  
neglecting recoil of the Brownian particle, and of treating the bath as 
a collection of free particles of mass $m$.  We  consider the pure state 
density matrix corresponding to definite momenta $\{\vec k_i\}$ of the 
bath particles, calculate the corresponding order $n$ density tensor defined 
by Eqs.~(32a-c), 
and then average over the thermal distribution of the bath particles 
as in Eqs.~(35a,b).  Thus the initial density matrix for the total system,  
corresponding to the factor $\rho_{\ell}$ in Eq.~(33d), is
$$\rho_{\ell}^{\rm TOT}=\rho_{\ell} \rho_{\cal E}~~~,\eqno(38a)$$
with $\rho_{\ell}$ the initial density matrix of the Brownian particle, 
characterized by its coordinate matrix elements $\langle \vec R_{\ell} 
|\rho_{\ell}| \vec R_{\ell}^{\,\prime} \rangle$, and with $\rho_{\cal E}$ the 
product density matrix for the bath particles, 
$$\rho_{\cal E}=\prod_i|\vec k_i \rangle \langle \vec k_i|  ~~~.\eqno(38b)$$.

Since the bath particle scatterings are all independent, we focus on the  
effect of the scattering of a single bath 
particle, of initial momentum $\vec k$, on the Brownian particle, which we   
take to be in a superposition of position eigenstates.  
Thus the initial state of the Brownian particle 
and the bath particle that we are considering is 
$$|I\rangle =\sum_{\vec R} c_{\vec R}|\vec R\,\rangle |\vec k\,\rangle 
~~~,\eqno(39a)$$ 
corresponding to an initial state density matrix 
$$\eqalign{
\rho_I=&|I\rangle \langle I|\cr
 =&\sum_{\vec R} \sum_{\vec R^{\prime}}c_{\vec R} c^*_{\vec R^{\prime}} 
|\vec R\,\rangle |\vec k\rangle  \langle \vec k|\langle \vec R^{\prime}|
~~~.\cr
}\eqno(39b)$$ 
The corresponding Brownian particle matrix element of $\rho_I$, which 
is still an operator on the bath particle state, takes the form  
$$\langle \vec R|\rho_I|\vec R^{\prime} \rangle 
=\rho(\vec R,\vec R^{\prime}) | \vec k\rangle \langle \vec k|
~~~,\eqno(39c)$$ 
with 
$$\rho(\vec R,\vec R^{\prime})= c_{\vec R} c^*_{\vec R^{\prime}}  ~~~.
\eqno(39d)$$ 

Asymptotically, the effect of the scattering is to replace the initial 
state $|I\rangle$ by $|F\rangle = S|I\rangle $, with $S$ the scattering 
matrix.  Substituting Eq.~(39a), and using translation invariance to 
relate the scattering matrix $S$ with the Brownian particle 
at a general coordinate, to the scattering matrix $S_0$ with the Brownian 
particle at the origin, we get [11]
$$\eqalign{
|F\rangle =&S|I\rangle 
=\sum_{\vec R} c_{\vec R}S|\vec R\,\rangle |\vec k\,\rangle \cr
=& \sum_{\vec R} c_{\vec R} |\vec R\, \rangle 
e^{-i\vec k_{\rm OP} \cdot \vec R } S_0 
e^{i\vec k_{\rm OP}  \cdot \vec R }   |\vec k\,\rangle~~~,\cr
}\eqno(40a)$$ 
with $\vec k_{\rm OP}$ the momentum operator for the bath particle.  
The corresponding final density matrix is then 
$$\eqalign{
\rho_F=& |F \rangle \langle F| \cr
 =&\sum_{\vec R} \sum_{\vec R^{\prime}}c_{\vec R} c^*_{\vec R^{\prime}} 
|\vec R\,\rangle 
e^{-i\vec k_{\rm OP} \cdot \vec R } S_0 
e^{i\vec k_{\rm OP}  \cdot \vec R }   |\vec k\,\rangle  \langle \vec k|
e^{-i\vec k_{\rm OP} \cdot \vec R^{\prime} } S_0^{\dagger} 
e^{i\vec k_{\rm OP}  \cdot \vec R^{\prime} }  
\langle \vec R^{\prime}| ~~~,\cr
}\eqno(40b)$$ 
and the  Brownian particle matrix element of $\rho_F$, which is again 
an operator acting on the bath particle, is 
$$
\langle \vec R|\rho_F|\vec R^{\prime} \rangle 
=\rho(\vec R,\vec R^{\prime})  
e^{-i\vec k_{\rm OP} \cdot \vec R } S_0 
e^{i\vec k_{\rm OP}  \cdot \vec R }   |\vec k\rangle  \langle \vec k|
e^{-i\vec k_{\rm OP} \cdot \vec R^{\prime} } S_0^{\dagger} 
e^{i\vec k_{\rm OP}  \cdot \vec R^{\prime} }~~~.  
\eqno(40c)$$
Substituting this expression into Eq.~(33d), we get 
$$\eqalign{
&\rho^{(n)}_{\vec R_1\vec R_1^{\prime},...,\vec R_n\vec R_n^{\prime};F} \cr
=&\prod_{\ell=1}^n \rho(\vec R_{\ell},\vec R_{\ell}^{\prime}) 
\langle \vec k|
e^{-i\vec k_{\rm OP} \cdot \vec R_{\ell}^{\prime} } S_0^{\dagger} 
e^{i\vec k_{\rm OP}  \cdot \vec R_{\ell}^{\prime} } 
e^{-i\vec k_{\rm OP} \cdot \vec R_{\ell+1} } S_0 
e^{i\vec k_{\rm OP}  \cdot \vec R_{\ell+1} }|\vec k \rangle  \cr 
=&\prod_{\ell=1}^n \rho(\vec R_{\ell},\vec R_{\ell}^{\prime}) 
\langle \vec k|
S_0^{\dagger} e^{i\vec k_{\rm OP}  \cdot 
(\vec R_{\ell}^{\prime}-\vec R_{\ell+1} )} S_0 |\vec k\rangle 
e^{i\vec k  \cdot (\vec R_{\ell+1}-\vec R_{\ell}^{\prime}) }~~~, \cr 
}\eqno(40d)$$
with $\vec R_{n+1}=\vec R_1$.  The matrix element appearing in the 
final line of Eq.~(40d) is one that is familiar from the standard 
calculation of the reduced density matrix (that is, $\rho^{(1)}_{\vec R, 
\vec R^{\prime}}$) for collisional decoherence [12].  Writing 
$$
\langle \vec k|
S_0^{\dagger} e^{i\vec k_{\rm OP}  \cdot 
(\vec R_{\ell}^{\prime}-\vec R_{\ell+1} )} S_0 |\vec k\rangle 
e^{i\vec k  \cdot (\vec R_{\ell+1}-\vec R_{\ell}^{\prime}) } 
=1+f(\vec R_{\ell+1}-\vec R_{\ell}^{\prime})~~~,\eqno(41a)$$
with $f$ proportional to the square of the scattering amplitude, 
the product of matrix elements in Eq.~(40d) can be written, to second 
order accuracy in the scattering amplitude, as
$$\prod_{\ell=1}^n [1+f(\vec R_{\ell+1}-\vec R_{\ell}^{\prime})] 
\simeq 1+ \sum_{\ell=1}^n f(\vec R_{\ell+1}-\vec R_{\ell}^{\prime})  
~~~.\eqno(41b)$$
We also note that Eq.~(39c), when substituted into Eq.~(32c), implies that  
the value of $\rho^{(n)}$ before the scattering is 
$$\rho^{(n)}_{\vec R_1\vec R_1^{\prime},...,\vec R_n\vec R_n^{\prime};I} 
=\prod_{\ell=1}^n \rho(\vec R_{\ell},\vec R_{\ell}^{\prime})~~~.\eqno(41c)$$ 
Thus when the approximation of Eq.~(41b) is substituted into Eq.~(40d), 
we get  
$$
\rho^{(n)}_{\vec R_1\vec R_1^{\prime},...,\vec R_n\vec R_n^{\prime};F}  
-\rho^{(n)}_{\vec R_1\vec R_1^{\prime},...,\vec R_n\vec R_n^{\prime};I} 
= [\sum_{\ell=1}^n f(\vec R_{\ell+1}-\vec R_{\ell}^{\prime}) ] 
\rho^{(n)}_{\vec R_1\vec R_1^{\prime},...,\vec R_n\vec R_n^{\prime};I} 
~~~.\eqno(41d)$$

At this point our work is essentially finished, since the remaining steps 
are identical to the standard calculation [11,12,13] proceeding from the $n=1$ case 
of Eq.~(41d), and the structure of Eq.~(41d) makes it clear how to generalize 
the standard result for $\rho^{(1)}$ to the case of general $\rho^{(n)}$. 
In brief, the standard procedure is to multiply the right hand side of 
Eq.~(41d) by the number of scattering particles, which combines with a 
normalizing factor of the inverse volume to give an overall factor of 
$N$, the scattering particle density.  The effect of the thermal distribution 
$\mu(\vec k)$ of momenta $\vec k$ is taken into account by 
including an integral $\int d\vec k \mu(\vec k)$, in accordance with 
the mixed state procedure of Eq.~(35a).  Finally, expressing the $S$ 
matrix in terms of the scattering amplitude $f(\vec k^{\prime},\vec k)$, 
and noting that the squared 
delta function for energy conservation  gives an overall factor of the 
elapsed time, Eq.~(41d) becomes, in the limit of small elapsed time, 
a formula for the time derivative 
of $\rho^{(n)}$.   For the $n=1$ case, the standard answer obtained this 
way is 
$$ {\partial \rho^{(1)}(t)_{\vec R\vec R^{\prime}} \over \partial t}
=-F(\vec R-\vec R^{\prime})  \rho^{(1)}(t)_{\vec R\vec R^{\prime}} ~~~,
\eqno(42a)$$
with 
$$F(\vec R)=N \int d\vec k \mu(\vec k) {|\vec k|\over m }
\int d\hat n  \big( 1-e^{i(\vec k - \hat n |\vec k|) \cdot \vec R)} \big) 
|f(\hat n |\vec k|,\vec k)|^2~~~,\eqno(42b)$$ 
where $\hat n$ is a unit vector which gives the direction of the 
scattered particle momentum $\vec k^{\prime}=\hat n |\vec k|$.  To compare Eq.~(42b) with 
the $n=1$ case of Eq.~(41d), we replace $\vec R$ by $ \vec R_2=\vec R_1$ and 
$\vec R^{\prime}$ by $\vec R^{\prime}_1$.  Then we see that the generalization 
to $n \geq 1$ is given by 
$${\partial \rho^{(n)}(t)_{\vec R_1\vec R_1^{\prime},...,
\vec R_n\vec R_n^{\prime}} \over \partial t}  
= - [\sum_{\ell=1}^n F(\vec R_{\ell+1}-\vec R_{\ell}^{\prime}) ] 
\rho^{(n)}(t)_{\vec R_1\vec R_1^{\prime},...,\vec R_n\vec R_n^{\prime}}. 
~~~\eqno(42c)$$

This is our final result for collisional Brownian motion, giving the 
evolution equation obeyed by the order $n$  density tensor .  
We see that it has the 
generic symmetries expected in the quantum noise case: although not totally 
symmetric in its pair indices, $\rho^{(n)}$ is symmetric under cyclic 
permutation of these indices.  As additional checks, we see that 
for $n=2$ the factor involving  $F$ is 
$$F(\vec R_2-\vec R_1^{\prime})+F(\vec R_1-\vec R_2^{\prime}) ~~~,\eqno
(43a)$$
which is symmetric under the interchange $1 \leftrightarrow 2$, 
while for $n=3$ we have 
$$ \eqalign{
F(\vec R_2-\vec R_1^{\prime})&+F(\vec R_3-\vec R_2^{\prime})
+F(\vec R_1-\vec R_3^{\prime}) =F^S + F^A~~~, \cr
F^S=& {1\over 2} [
F(\vec R_2-\vec R_1^{\prime})+F(\vec R_3-\vec R_2^{\prime})
+F(\vec R_1-\vec R_3^{\prime}) \cr 
+&F(\vec R_1-\vec R_2^{\prime})+F(\vec R_3-\vec R_1^{\prime})
+F(\vec R_2-\vec R_3^{\prime})  ]  ~~~,\cr
F^A=& {1\over 2} [
F(\vec R_2-\vec R_1^{\prime})+F(\vec R_3-\vec R_2^{\prime})
+F(\vec R_1-\vec R_3^{\prime}) \cr 
-&F(\vec R_1-\vec R_2^{\prime})-F(\vec R_3-\vec R_1^{\prime})
-F(\vec R_2-\vec R_3^{\prime})  ]  ~~~,\cr
}\eqno(43b)$$ 
with $F^S$ symmetric, and $F^A$ antisymmetric, under any of the pair interchanges 
$1 \leftrightarrow 2$, or $1 \leftrightarrow 3$, or $2 \leftrightarrow 3$.  
Checking the descent equations is easy. Setting $\vec R_1^{\prime}=\vec R_2$, 
the term $F(\vec R_2-\vec R_1^{\prime})$ in Eq.~(42c) vanishes, so that 
on integrating over $\vec R_1^{\prime}$ one is left on the right hand side 
with a sum  $F(\vec R_1-\vec R_n^{\prime})
+F(\vec R_3-\vec R_2^{\prime})+...$ 
that does not involve $\vec R_1^{\prime}$, times 
$$\int d\vec R_1^{\prime} 
\rho^{(n)}(t)_{\vec R_1\vec R_1^{\prime},\vec R_1^{\prime}\vec R_2^{\prime},
...,\vec R_n\vec R_n^{\prime}}~~~,\eqno(43c)$$  
and so the descent equation for 
$\rho^{(n)}(t)$ then implies the descent equation for its time derivative. 
\bigskip
\vfill\eject

\centerline{\bf 8.~~The weak coupling Born-Markov approximation and 
the quantum optical }
\centerline{\bf  master equation for the density tensor}
\bigskip
We turn next to the density tensor extension of the standard weak coupling 
Born-Markov approximation, that is used to give a master equation for the 
reduced density matrix $\rho^{(1)}$ for a system ${\cal S}$ interacting with 
an environment ${\cal E}$. We assume a total system plus 
environment Hamiltonian $H_{\rm TOT}=H_{\cal E} + H_{\cal S} + H$, 
with $H_{\cal E}$ and $H_{\cal S}$ 
respectively the environment and system Hamiltonians, and with    
$H$ the system-environment interaction Hamiltonian.  (We omit the customary  subscript 
$I$ on the interaction Hamiltonian to avoid a proliferation of subscripts.)
We shall work in this section in interaction picture, in which the operators 
carry the time dependence associated with $H_{\cal E}$ and $H_{\cal S}$.  
Thus the interaction Hamiltonian carries a time dependence $H(t)$, 
and the density matrix obeys the equation of motion 
$${d \rho(t) \over dt}=-i[H(t), \rho(t)]~~~\eqno(44a)$$ 
which can be integrated to give 
$$\rho(t)=\rho(0)-i\int_0^t ds [H(s),\rho(s)]~~~.\eqno(44b)$$ 
Substituting Eq.~(44b) back into Eq.~(44a) gives the additional 
evolution equation 
$${d \rho(t) \over dt}=-i[H(t), \rho(0)] - 
\int_0^t ds[H(t), [H(s),\rho(s)]]~~~.\eqno(44c)$$ 
One then notes that up to an error of order $H^3$, the time argument 
of the factor $\rho(s)$ in the double commutator term is irrelevant, 
so this factor can be approximated as $\rho(t)$, 
giving 
$${d \rho(t) \over dt}=-i[H(t), \rho(0)] - 
\int_0^t ds[H(t), [H(s),\rho(t)]]~~~,\eqno(44d)$$ 
which is used as the starting point for the standard master equation 
derivation.  

Our first step is to derive a suitable  extension of Eq.~(44d) for 
the product $\rho_1\rho_2...\rho_n$ that appears in Eq.~(32c).  By the chain 
rule, we have 
$${d (\rho_1\rho_2...\rho_n) \over dt}= {d \rho_1 \over dt} \rho_2...\rho_n 
+\rho_1...{d\rho_{\ell} \over dt}...\rho_n + \rho_1....{d\rho_n\over dt}~~~.
\eqno(45a)$$  
For each undifferentiated factor on the right of Eq.~(45a) we substitute 
Eq.~(44b), and for each time derivative factor we substitute Eq.~(44c), 
with appropriate subscripts added.  Let us now organize the terms  
obtained this way according to the number of factors of $H$ that appear. 
Since Eq.~(44c) contains at least one factor of $H$, there are 
no terms in Eq.~(45a) with no factors of $H$.  The general term in Eq.~(45a)
with one factor of $H$ comes from 
the term in Eq.~(44c) with one factor of $H$, multiplied by the product 
of the terms from Eq.~(44b) with no factors of $H$, giving 
$$-i\big( [H_1(t),\rho_1(0)] \rho_2(0)...\rho_n(0) + \rho_1(0) 
[H_2(t),\rho_2(0)]...\rho_n(0)+...+\rho_1(0)\rho_2(0)...[H_n(t),\rho_n(0)] 
\big) ~~~.\eqno(45b)$$
The terms in Eq.~(45a) with two factors of $H$ are of 
two types: (1) the quadratic term 
in $H$ 
on the right of Eq.~(44c) times factors of $\rho(0)$, and (2) the linear 
term in $H$ on the right of Eq.~(44c), multiplied by  one factor of the linear 
term on the right of Eq.~(44b), times factors of $\rho(0)$.  We now note 
that up to an error of order $H^3$, in terms that already contain two 
factors of $H$ we can replace all factors $\rho(0)$ or $\rho(s)$ by the 
corresponding $\rho(t)$, since the differences $\rho(t)-\rho(s)$ and 
$\rho(t)-\rho(0)$ are all of order $H$.  Collecting everything, we get 
the following formula, which gives the needed extension of Eq.~(44d), 
$$\eqalign{
&{d (\rho_1\rho_2...\rho_n) \over dt}
=-i\sum_{\ell=1}^n 
\rho_1(0)...\rho_{\ell-1}(0) [H_{\ell}(t),\rho_{\ell}(0)]\rho_{\ell+1}(0) 
...\rho_n(0)\cr 
&-\sum_{\ell=1}^n 
\rho_1(t)...\rho_{\ell-1}(t)
\int_0^t ds  [H_{\ell}(t),[H_{\ell}(s),\rho_{\ell}(t)]]
\rho_{\ell+1}(t) ...\rho_n(t)\cr 
&-\sum_{\ell<m} \{  \rho_1(t)...\rho_{\ell-1}(t) 
[H_{\ell}(t),\rho_{\ell}(t)]
\rho_{\ell+1}(t)...\rho_{m-1}(t)\int_0^t ds [H_m(s),\rho_m(t)] 
\rho_{m+1}(t)...\rho_n(t)\cr 
&+\rho_1(t)...\rho_{\ell-1}(t)\int_0^t ds [H_{\ell}(s),\rho_{\ell}(t)]
\rho_{\ell+1}(t)...\rho_{m-1}(t) [H_m(t),\rho_m(t)] 
\rho_{m+1}(t)...\rho_n(t)\}  +{\rm O}(H^3)  ~~~.\cr
}\eqno(45c)$$
Taking the overall ${\rm Tr}_{\cal E}$ of this expression then gives 
a formula for the time evolution of $\rho^{(n)}(t)$ as defined by 
Eq.(32c). 

We now make two standard assumptions.  First of all, we assume 
at that at the initial time $t=0$, the density matrix factorizes so  
that $\rho(0)= \rho_{\cal E} \rho_{\cal S} $, with $\rho_{\cal E}$ 
and $\rho_{\cal S}$ respectively density matrices for 
the environment and the system  which commute with one another, and with 
$\rho_{\cal E}$ a pure state density matrix obeying 
$\rho_{\cal E}^2=\rho_{\cal E}$.  
Secondly, we assume that  $\langle H \rangle_{\cal E}= 
{\rm Tr}_{\cal E} \rho_{\cal E} H
=0$, that is, we take the interaction Hamiltonian to have a vanishing 
expectation in the initial environmental state.  As a result of these 
two assumptions, the environmental trace of the first term on 
the right hand side of Eq.~(45c) vanishes, since  
$$\eqalign{
&{\rm Tr}_{\cal E} 
\rho_1(0)...\rho_{\ell-1}(0) [H_{\ell}(t),\rho_{\ell}(0)]\rho_{\ell+1}(0) 
...\rho_n(0) \cr
=& \rho_{{\cal S}1}...\rho_{{\cal S}\ell-1}
[({\rm Tr}_{\cal E} \rho_{\cal E} H_{\ell}(t)),\rho_{{\cal S}\ell}]
\rho_{{\cal S}\ell+1} 
...\rho_{{\cal S}n}=0~~~.\cr 
}\eqno(46a)$$ 
The remaining terms in Eq.~(45c) all have two factors of $H$.  Since  
$\rho(t)$ and $\rho(0)$ differ by one power of $H$, in these 
terms, up to an error of order $H^3$, we can replace all factors $\rho(t)$ 
by the factorized approximation 
$$\rho(t)\simeq \rho(0)=\rho_{\cal E}  \rho_{\cal S} 
= \rho_{\cal E} {\rm Tr}_{\cal E}  \rho(0)  
\simeq  \rho_{\cal E} {\rm Tr}_{\cal E}  \rho(t)  
=\rho_{\cal E} \rho^{(1)}(t)
~~~.\eqno(46b)$$
With these simplifications, and remembering that system operator factors 
$\rho_{\ell}^{(1)}$ with different index values $\ell$ act on different 
Hilbert spaces ${\cal H}_{{\cal S};\ell}$ and so commute, Eq.~(45c) 
becomes an extended version of the Redfield equation, 
$$\eqalign{
&d\rho^{(n)}(t)/dt =
-\sum_{\ell=1}^n (\rho^{(1)}_1(t)...\rho^{(1)}_n(t))_{\ell}  
{\rm Tr}_{\cal E} \rho_{\cal E}^{n-1} \int_0^t ds 
[H_{\ell}(t),[H_{\ell}(s),\rho_{\ell}^{(1)}(t)\rho_{\cal E}]] \cr
&-\sum_{\ell< m}  (\rho^{(1)}_1(t)...\rho^{(1)}_n(t))_{\ell m} 
{\rm Tr}_{\cal E} \int_0^t ds \rho_{\cal E}^{n-(m-\ell)-1} \cr
&\times \{[ H_{\ell}(t), \rho^{(1)}_{\ell}(t) \rho_{\cal E}] 
\rho_{\cal E}^{m-\ell-1} [H_m(s), \rho_m^{(1)}(t) \rho_{\cal E}]
+[ H_{\ell}(s), \rho^{(1)}_{\ell}(t) \rho_{\cal E}] 
\rho_{\cal E}^{m-\ell-1} [H_m(t), \rho_m^{(1)}(t) \rho_{\cal E}] \}
~~~.  \cr
}\eqno(46c)$$
This is converted to the Born-Markov equation by setting $s \to t-s$, 
and then extending the upper limit of the $s$ integration from $t$ to 
$\infty$, giving 
$$\eqalign{
&d\rho^{(n)}(t)/dt =
-\sum_{\ell=1}^n (\rho^{(1)}_1(t)...\rho^{(1)}_n(t))_{\ell} 
{\rm Tr}_{\cal E} \rho_{\cal E}^{n-1} \int_0^{\infty} ds 
[H_{\ell}(t),[H_{\ell}(t-s),\rho_{\ell}^{(1)}(t)\rho_{\cal E}]] \cr
&-\sum_{\ell< m}  (\rho^{(1)}_1(t)...\rho^{(1)}_n(t))_{\ell m}  
{\rm Tr}_{\cal E} \int_0^{\infty} ds \rho_{\cal E}^{n-(m-\ell)-1} \cr
&\times \{[ H_{\ell}(t), \rho^{(1)}_{\ell}(t) \rho_{\cal E}] 
\rho_{\cal E}^{m-\ell-1} [H_m(t-s), \rho_m^{(1)}(t) \rho_{\cal E}]
+[ H_{\ell}(t-s), \rho^{(1)}_{\ell}(t) \rho_{\cal E}] 
\rho_{\cal E}^{m-\ell-1} [H_m(t), \rho_m^{(1)}(t) \rho_{\cal E}] \}
~~~.  \cr
}\eqno(46d)$$

We now note that Eq.~(46d) can be further simplified, by taking account 
of the fact that whenever an $H$ factor is sandwiched between factors 
of $\rho_{\cal E}$ it vanishes, since $\rho_{\cal E} H \rho_{\cal E} 
=\rho_{\cal E} \langle H \rangle_{\cal E}=0$.  
This eliminates all terms in the 
sum over $\ell, m$ that are not adjacent in a cyclic sense, i.e., that 
do not either have $m=\ell+1$, $\ell=1,...,n-1$, or $\ell=1, m=n$.  The 
latter, by use of the cyclic properties of the trace, can be rearranged 
to give the $\ell=n$ term of the former set.  
We thus get a simplified set of Born-Markov equations.  
For $n=1$, we 
get the usual starting  point for the Born-Markov master equation 
derivation, 
$$\eqalign{
d\rho^{(1)}(t)/dt=-{\rm Tr}_{\cal E} \int_0^{\infty}ds 
&[H(t)H(t-s)\rho^{(1)}(t)\rho_{\cal E}+\rho^{(1)}(t) \rho_{\cal E} 
H(t-s) H(t)  \cr
-&H(t)\rho^{(1)}(t)\rho_{\cal E} H(t-s)-H(t-s)\rho^{(1)}(t) \rho_{\cal E}
H(t)]~~~,\cr
}\eqno(47a)$$ 
and for $n\geq 2$, with the subscript $n+1$ identified with 1, 
$$\eqalign{
&d\rho^{(n)}(t)/dt=-{\rm Tr}_{\cal E}  \rho_{\cal E} \int_0^{\infty} ds \sum_{\ell=1}^n \cr
&\times \{  (\rho^{(1)}_1(t)...\rho^{(1)}_n(t))_{\ell} 
[H_{\ell}(t)H_{\ell}(t-s) \rho_{\ell}^{(1)}(t) 
+ \rho_{\ell}^{(1)}(t) H_{\ell}(t-s) H_{\ell}(t) ]\cr 
&-  (\rho^{(1)}_1(t)...\rho^{(1)}_n(t))_{\ell \ell+1}  
[\rho^{(1)}_{\ell}(t) H_{\ell}(t) H_{\ell+1}(t-s)\rho^{(1)}_{\ell+1}(t)   
+\rho^{(1)}_{\ell}(t) H_{\ell}(t-s) 
H_{\ell+1}(t)\rho^{(1)}_{\ell+1}(t)]\} ~~~.\cr   
}\eqno(47b)$$
At this point it is useful to check (and we have done so) that the 
descent equations are satisfied by Eqs.~(47a) and (47b). 

The remainder of the derivation follows closely the standard master 
equation derivation, in the rotating wave approximation, that proceeds 
from Eq.~(47a), so we will only give a sketch. For further details,  
and in particular a discussion of the physical 
justification for the approximations 
involved, see Sec. 3.3 of ref [1] and also ref [13].  One assumes that 
$H_{\ell}(t)$ has the form 
$$H_{\ell}(t)=\sum_{\alpha}\sum_{\omega} e^{i\omega t} A_{\ell \alpha}^{\dagger} 
(\omega) B_{\alpha}(t)~~~,\eqno(48a)$$
with $A_{\ell \alpha}^{\dagger}$ acting only in the system Hilbert 
space ${\cal H}_{{\cal S}; \ell}$ 
and with $B_{\alpha}$  acting only in the environment Hilbert space 
${\cal H}_{\cal E}$, and with the Hermiticity properties 
$A_{\ell \alpha}^{\dagger}(\omega)=A_{\ell \alpha}(-\omega)$ and 
$B_{\alpha}^{\dagger}(t)=B_{\alpha}(t)$.  Since Eqs.~(47a,b) are quadratic in 
$H$, one uses Eq.~(48a) twice; for each $H_{k}(t-s)$ (regardless of the 
value of the index $k$) one writes 
$$H_k(t-s)=\sum_{\beta \omega} e^{-i\omega(t-s)} A_{k\beta}(\omega)
B_{\beta}(t-s)~~~,\eqno(48b)$$ 
and for each $H_k(t)$ (again regardless of the value of $k$) one writes 
$$H_{k}(t)=\sum_{\alpha\omega^{\prime}} e^{i\omega^{\prime}t} 
A_{k\alpha}^{\dagger}(\omega^{\prime})B_{\alpha}^{\dagger}(t)~~~.\eqno(48c)$$
The rotating wave approximation then consists of neglecting terms in 
the double sum with $\omega^{\prime} \not= \omega$, so that only the 
diagonal terms $\omega^{\prime}=\omega$ are left.  
From the trace over the environment, and the integral over $s$, one 
gets correlators of the form 
$$\eqalign{
&\int_0^{\infty} ds e^{i\omega s}\langle B_{\alpha}^{\dagger}(t)   
B_{\beta}(t-s) \rangle_{\cal E}\equiv \Gamma_{\alpha\beta}(\omega)~~~,\cr  
&\int_0^{\infty} ds e^{i\omega s} \langle B_{\beta}(t-s) 
B_{\alpha}^{\dagger}(t) \rangle_{\cal E}= \Gamma_{\alpha\beta}(-\omega)^*
~~~,\cr}\eqno(49a)$$ 
where in the second line we have used the definition of the first line and
the adjointness properties of the integrand.  It is also customary to 
decompose the reservoir correlation function $\Gamma_{\alpha\beta}$ into self-adjoint 
and anti-self-adjoint parts, according to 
$$\Gamma_{\alpha \beta}(\omega)={1\over 2} \gamma_{\alpha\beta}(\omega) 
+i S_{\alpha\beta}(\omega)~~~.\eqno(49b)$$ 

Proceeding in this fashion, after some algebra one gets the final result, 
which can be written as an equation for all $n \geq 1$ 
by including a $\delta_{n1}$ to take account of the special nature 
of the $n=1$ equation,  
$$\eqalign{
d\rho^{(n)}(t)/dt =&  \sum_{\ell=1}^n 
(\rho^{(1)}_1(t)...\rho^{(1)}_n(t))_{\ell} 
i[\rho^{(1)}_{\ell}(t), \sum_{\omega \alpha \beta} 
S_{\alpha \beta}(\omega) A_{\ell \alpha}^{\dagger}(\omega) 
A_{\ell \beta}(\omega)] \cr
+& \sum_{\ell=1}^n \sum_{\omega\alpha\beta}\gamma_{\alpha \beta}(\omega) 
\left\{(\rho^{(1)}_1(t)...\rho^{(1)}_n(t))_{\ell}
\left[\delta_{n1} A_{\ell \beta}(\omega) \rho^{(1)}_{\ell}(t)  
A_{\ell \alpha}^{\dagger} (\omega) 
-{1\over 2}\{ A_{\ell \alpha}^{\dagger} (\omega)A_{\ell \beta}(\omega),  
\rho^{(1)}_{\ell}(t) \} \right] \right.\cr 
+&\left.  (\rho^{(1)}_1(t)...\rho^{(1)}_n(t))_{\ell \ell+1} 
 \rho^{(1)}_{\ell}(t)   
 A_{\ell \alpha}^{\dagger}(\omega) A_{\ell+1 \beta}(\omega)
 \rho^{(1)}_{\ell+1}(t)  \right\}~~~.
 }\eqno(50a)$$
Despite the fact the the $n=1$ and $n\geq2$ density tensors have 
a different structure, the descent equations are satisfied by Eq.~(50a), 
as verified in Appendix D.  

Finally, we note that Eq.~(50a) is readily converted to the quantum 
optical master equation and its density tensor generalizations, by 
taking $\alpha$ to be a three-vector index, so that $A_{\alpha}$ becomes 
$\vec A$, which is related to the dipole operator by Eq.~(3.182) of  
ref [1].  Also, one takes 
$S_{\alpha \beta}(\omega)=\delta_{\alpha\beta}S(\omega)$,  with 
$S(\omega)$ given by Eq.~(3.205) of ref [1], and $\gamma_{\alpha \beta}
(\omega)=(4\omega^3/3) [1+N(\omega)]\delta_{\alpha \beta}$, with 
$N(\omega)=1/(e^{\beta\omega}-1)$ the photon number operator.  
One gets in this way the density tensor generalization of the   
quantum optical master equation, 
$$\eqalign{
&d\rho^{(n)}(t)/dt =  \sum_{\ell=1}^n 
(\rho^{(1)}_1(t)...\rho^{(1)}_n(t))_{\ell} 
i[\rho^{(1)}_{\ell}(t), \sum_{\omega} 
S(\omega) \vec A_{\ell }^{\,\dagger}(\omega)  \cdot 
\vec A_{\ell}(\omega)] \cr
&+   \sum_{\ell=1}^n \sum_{\omega}(4\omega^3/3)[1+N(\omega)] 
\left\{(\rho^{(1)}_1(t)...\rho^{(1)}_n(t))_{\ell}
\left[\delta_{n1} \vec A_{\ell}(\omega) \cdot \rho^{(1)}_{\ell}(t)  
\vec A_{\ell}^{\,\dagger} (\omega) 
-{1\over 2}\{ \vec A_{\ell }^{\,\dagger}(\omega) \cdot \vec A_{\ell }(\omega),  
\rho^{(1)}_{\ell}(t) \} \right] \right.\cr 
&+\left.  (\rho^{(1)}_1(t)...\rho^{(1)}_n(t))_{\ell \ell+1} 
 \rho^{(1)}_{\ell}(t)   
\vec  A_{\ell }^{\,\dagger}(\omega)\cdot \vec A_{\ell+1 }(\omega)
 \rho^{(1)}_{\ell+1}(t)  \right\}~~~,\cr
 }\eqno(50b)$$
which is our final result of this section.  
\bigskip
\centerline{\bf 9.~~The Caldeira--Leggett model master equation for the  
density tensor}
\bigskip
The Caldeira--Leggett model [14] describes the damping of the one-dimensional 
motion of a Brownian particle of mass m, moving in a potential $V(x)$, 
and interacting with an environment consisting of harmonic oscillators 
with masses $m_o$ and frequencies $\omega_o$, and annihilation operator   
$b_o$.  The interaction Hamiltonian 
is assumed to be a linear coupling $H=-xB$, with 
$$B=\sum_o\kappa_o x_o=\sum_o \kappa_o (b_o + b_o^{\dagger})/(2m_o \omega_o)
^{1\over 2}~~~\eqno(51a)$$ 
a weighted sum of the harmonic oscillator coordinates.  A counter-term 
formally of order $H^2$, 
$$H_c=x^2\sum_o {\kappa_o^2\over 2 m_o \omega_o}\equiv x^2 C~~~,\eqno(51b)$$ 
is included in the 
calculation, so that the total Hamiltonian is 
$$H_{\rm TOT}=H_{\cal E} + H_{\cal S} + H + H_c~~~,\eqno(52a)$$ 
with $H_{\cal E}$ and $H_{\cal S}$ respectively the oscillator and 
particle Hamiltonians, 
$$\eqalign{
H_{\cal E}=&\sum_o \omega_o (b_o^{\dagger}b_o +{1\over 2})~~~,\cr
H_{\cal S}=&{p^2\over 2m} +V(x)~~~.\cr 
}\eqno(52b)$$ 
Our aim will be to get a description of the effect on the particle motion 
of the couplings to the oscillator environment, in the high temperature  
limit.  
Our derivation of the density tensor generalization of the high temperature 
master equation closely follows that of Sec. 3.6 of ref [1], to which 
the reader is referred for a discussion of the physical motivation of 
the approximations involved.  

Since the environmental expectation of the interaction Hamiltonian $H$ 
vanishes, we can proceed directly from the simplified Born-Markov 
equation of Eqs.~(47a) and (47b).  The first step is to transform the  
density matrix $\rho(t)$ 
back to Schr\"odinger picture; it is easy to see that the effect of 
this is to replace $H(t)$ by $H=H(0)$, to replace $H(t-s)$ by $H(-s)$ 
(with $H(-s)$ still in the interaction picture), and to change $d/dt$ to 
$D/dt$, defined by 
$$D\rho^{(n)}(t)/dt = d\rho^{(n)}(t)/dt + i \sum_{\ell=1}^n 
{\rm Tr}_{\cal E} \rho_1(t)...\rho_{\ell-1}(t)[p_{\ell}^2/(2m)+V(x_{\ell}),
\rho_{\ell}(t)] \rho_{\ell+1}(t)...\rho_n(t) ~~~.\eqno(53a)$$
It is also necessary to explicitly include commutators arising from the 
counter term, which is easy since this term is treated as being already 
quadratic in $H$.  For the analog of Eq.~(47a) for the special case 
$n=1$, we find 
$$\eqalign{
D\rho^{(1)}(t)/dt= -i[H_c,\rho^{(1)}(t) ]
-{\rm Tr}_{\cal E} \int_0^{\infty}ds 
&[H H(-s)\rho^{(1)}(t)\rho_{\cal E}+\rho^{(1)}(t) \rho_{\cal E} 
H(-s) H  \cr
-&H\rho^{(1)}(t)\rho_{\cal E} H(-s)-H(-s)\rho^{(1)}(t) \rho_{\cal E}H]~~~,\cr
}\eqno(53b)$$ 
and for the analog of Eq.~(47b) for $n\geq 2$, we have 
$$\eqalign{
d\rho^{(n)}(t)/dt=&-i\sum_{\ell=1}^n
(\rho^{(1)}_1(t)...\rho^{(1)}_n(t))_{\ell} 
[H_{c\ell},\rho^{(1)}_{\ell}(t)]\cr
-&{\rm Tr}_{\cal E} \rho_{\cal E} \int_0^{\infty} ds \sum_{\ell=1}^n \cr
\times& \{  (\rho^{(1)}_1(t)...\rho^{(1)}_n(t))_{\ell} 
[H_{\ell}H_{\ell}(-s) \rho_{\ell}^{(1)}(t) 
+ \rho_{\ell}^{(1)}(t) H_{\ell}(-s) H_{\ell} ]\cr 
-&  (\rho^{(1)}_1(t)...\rho^{(1)}_n(t))_{\ell \ell+1}  
[\rho^{(1)}_{\ell}(t) H_{\ell} H_{\ell+1}(-s)\rho^{(1)}_{\ell+1}(t)   
+\rho^{(1)}_{\ell}(t) H_{\ell}(-s) 
H_{\ell+1}\rho^{(1)}_{\ell+1}(t)]\} ~~~.\cr   
}\eqno(53c)$$
We next note that 
$$H_{\ell}=-x_{\ell}(0)B(0)~,~~H_{\ell}(-s)=-x_{\ell}(-s)B(-s)~~~,
\eqno(54a)$$
where, using the assumption that the system evolution is slow compared to 
the oscillator time scale, we approximate $x_{\ell}(-s)$ by its free particle 
dynamics, 
$$x_{\ell}(-s) \simeq x_{\ell}-{p_{\ell}\over m} s~~~.\eqno(54b)$$
Since the right hand sides of Eqs.~(53b,c) are quadratic in $H$, the operator $B$ giving the 
coupling to the oscillators appears, after the environmental trace is 
taken, only through the correlators 
$$\eqalign{
D(s) \equiv & i\langle [B(0),B(-s)]\rangle_{\cal E}~~~,\cr 
D_1(s) \equiv & \langle \{B(0),B(-s)\} \rangle _{\cal E}~~~,\cr
}\eqno(55a)$$
so that we have 
$$\eqalign{
\langle B(0)B(-s)\rangle_{\cal E}=&{1\over 2}[D_1(s)-iD(s)]~~~,\cr 
\langle B(-s)B(0)\rangle_{\cal E}=&{1\over 2}[D_1(s)+iD(s)]~~~.\cr 
}\eqno(55b)$$
These correlators appear in the following integrals, which are evaluated 
or approximated in Sec. 3.6.2 of ref [1], 
$$\eqalign{
\int_0^{\infty} ds D(s) =& 2C~~~,\cr 
\int_0^{\infty} ds D_1(s)=& 4m\gamma k_B T~~~,\cr
\int_0^{\infty} ds s D(s) =&2m\gamma~~~,\cr
\int_0^{\infty} ds s D_1(s) =& 4m\gamma k_B T/\Omega \simeq 0~~~,\cr
}\eqno(55c)$$
with $C$ the constant defined by the counter term of Eq.~(51b), with 
$\gamma$ a constant determined by the harmonic oscillator spectral density, 
with $k_B$ and $T$ respectively the Boltzmann constant and environment 
temperature, and with $\Omega$ a frequency cutoff.  For a spectral 
density  $J(\omega)$  with a Lorentz-Drude cutoff function, one has 
$$J(\omega)=\sum_o {\kappa_o^2\over 2m_o\omega_o} \delta(\omega-\omega_o)
={2 m \gamma \over \pi} \omega {\Omega^2 \over \Omega^2 + \omega^2}~~~.\eqno(55d)$$

This completes the specification of the calculation; the rest is just the 
algebra of assembling all the pieces, and so we pass directly to the result. 
For $n=1$, we get the Caldeira--Leggett master equation, 
$$D\rho^{(1)}(t)/dt=-i\gamma [x,\{p,\rho^{(1)}(t)\}]
-2m\gamma k_B T [x,[x,\rho^{(1)}(t)]]~~~.\eqno(56a)$$
For the density tensors with $n\geq 2$, we correspondingly get 
$$\eqalign{
D\rho^{(n)}/dt=&\sum_{\ell=1}^n  (\rho^{(1)}_1(t)...\rho^{(1)}_n(t))_{\ell} 
 \left[ -2m\gamma k_B T \{ x_{\ell}^2,\rho^{(1)}_{\ell}(t) \} 
 + i\gamma\big( \rho^{(1)}_{\ell}(t) p_{\ell} x_{\ell}
 -x_{\ell} p_{\ell} \rho^{(1)}_{\ell}(t)  \big) \right]  \cr 
 +& \sum_{\ell=1}^n  (\rho^{(1)}_1(t)...\rho^{(1)}_n(t))_{\ell \ell+1} 
 [4m\gamma k_B T \rho^{(1)}_{\ell}(t) x_{\ell} x_{\ell+1} 
 \rho^{(1)}_{\ell+1}(t)
 +i\gamma \rho^{(1)}_{\ell}(t) (x_{\ell} p_{\ell+1}-p_{\ell} x_{\ell+1}) 
 \rho^{(1)}_{\ell+1}(t) ]~~~. \cr
 }\eqno(56b)$$
We also note that the term proportional to $\gamma$ on the first line of 
Eq.~(56b) can be written in the alternative form,
$$ i\gamma\big( \rho^{(1)}_{\ell}(t) p_{\ell} x_{\ell}
 -x_{\ell} p_{\ell} \rho^{(1)}_{\ell}(t)  \big) 
 =\gamma \rho^{(1)}_{\ell}(t) + {i\over 2} \gamma 
 [\rho^{(1)}_{\ell}(t),\{x_{\ell},p_{\ell} \} ]~~~.\eqno(56c)$$
Equations (56a) and (56b) are our final results for the 
Caldeira--Leggett model.  As was the case for the master equations derived in the preceding 
section, despite the differences between the structure of the $n=1$ and 
the $n \geq 2$ equations, the descent equations are satisfied, as verified 
in Appendix E.
\bigskip
\centerline{\bf 10.~~An application to state vector reduction}
\bigskip
We turn now to considerations that bridge the discussions given 
above in the classical and quantum noise cases.  We begin with an 
analysis of two It\^o stochastic Schr\"odinger equations, 
$$d|\psi\rangle =-{1\over 2} (A-\langle A\rangle)^2 dt |\psi\rangle  
 +(A-\langle A\rangle) dW_t  |\psi\rangle ~~~,\eqno(57a)$$ 
 and 
$$d|\psi\rangle =-{1\over 2} A^2 dt |\psi\rangle  
 +iA dW_t  |\psi\rangle ~~~,\eqno(57b)$$ 
with $dW_t$ a real Brownian noise obeying $dW_t^2=dt$, and where we have   
dropped the Hamiltonian term. These  lead 
to the respective  density matrix  evolution equations
$$d\rho=-{1\over 2} [A,[A,\rho]] dt + [\rho,[\rho,A]]dW_t~~~,\eqno(58a)$$
and 
$$d\rho=-{1\over 2} [A,[A,\rho]] dt + i[A,\rho]dW_t~~~,\eqno(58b)$$
which correspond to the same Lindblad type evolution equation 
for the expectation $E[\rho]$,
$$dE[\rho]={\cal L}E[\rho] dt~,~~{\cal L}\rho= -{1\over 2} [A,[A,\rho]]
~~~.\eqno(58c)$$

Let us now consider the effect of the stochastic 
evolutions of Eqs.~(58a,b,c) 
on the expectation of the variance $V={\rm Var}(A)$ of the operator $A$,   
$$\eqalign{
V =&{\rm Tr}\rho A^2 - ({\rm Tr} \rho A )^2~~~,\cr
E[V]=&{\rm Tr} E[\rho] A^2 - E[\rho_{i_1j_1}\rho_{i_2j_2}] 
A_{j_1i_1}A_{j_2i_2} \cr 
=& {\rm Tr} \rho^{(1)} A^2-\rho^{(2)}_{i_1j_1,i_2j_2}   
A_{j_1i_1}A_{j_2i_2}~~~, \cr 
}\eqno(59a)$$ 
where in the final line we have used the density tensor definition 
of Eq.~(15).  
For the time evolution of $E[V]$ we have  
$$\eqalign{
dE[V]/dt=&{\rm Tr}( {\cal L}E[\rho]) A^2  
-d\rho^{(2)}_{i_1j_1,i_2j_2} A_{j_1i_1}A_{j_2i_2}  \cr
=&{\rm Tr}( {\cal L}E[\rho]) A^2   
-2E[\rho_{i_1j_1} ({\cal L}\rho)_{i_2j_2}] A_{j_1i_1}   A_{j_2i_2}
-E[C_{i_1j_1,i_2j_2}] A_{j_1i_1}A_{j_2i_2} ~~~,\cr
}\eqno(59b)$$
where we have used Eq.~(58c) in the first line and Eq.~(23a) in the 
second line.   
Since the cyclic property of the trace implies that
${\rm Tr} [A,[A,\rho]]A=0$ , the terms in Eq.~(59b) involving the Lindblad
${\cal L}$  all vanish, and so the time derivative of $E[V]$ comes 
entirely form the final term, 
$$dE[V]/dt= -E[C_{i_1j_1,i_2j_2}] A_{j_1i_1}A_{j_2i_2} ~~~,\eqno(59c)$$ 
and thus is determined by the evolution equation for the second order 
density tensor.  This is why the state vector evolutions of Eqs.~(57a) and 
(57b), or equivalently the density matrix evolutions of 
Eqs.~(58a) and (58b), lead to very different results for the evolution 
of the variance of the operator $A$.  The tensor $C_{i_1j_1,i_2j_2}$ 
corresponding to Eqs.~(57b) and (58b) is given in Eq.~(24b), and since 
the cyclic property of the trace implies that ${\rm Tr} [A,\rho]A=0$,  
one has  $dE[V]/dt=0$ for this evolution.  On the other hand, the 
tensor $C_{i_1j_1,i_2j_2}$ corresponding to Eqs.~(57a) and (58a) is 
given in Eq.~(24a), and through Eq.~(59c) implies that 
$$dE[V]/dt=-E[ ({\rm Tr}[\rho,[\rho,A]]A )^2]
=-E[({\rm Tr} ([\rho,A])^2)^2] ~~~,\eqno(59d)$$
which is negative definite.  Starting from Eq.~(59d), some simple 
inequalities imply  that the stochastic evolution of Eqs.~(57a) and (58a) 
drives the variance of $A$ to zero as $t \to \infty$, and hence reduces 
the state vector to an eigenstate of $A$, as discussed in detail in  
refs [15]. 

Let us now consider a quantum system ${\cal S}$, consisting of a microscopic 
system coupled to a macroscopic measuring apparatus,  interacting 
with a quantum environment ${\cal E}$, with the totality forming a closed 
system.  A general result [16], using just the linearity of quantum mechanics, 
shows that state vector reduction cannot occur in this case.  To understand 
this result through an analysis similar to that just given 
for Eqs.~(57a,b), let us consider the behavior of the variance of a 
system operator  $A$ which is  a good ``pointer observable''.   By  
definition, a system operator commutes  with 
the environment Hamiltonian $H_{\cal E}$, and since the system in this case 
includes the apparatus and so is macroscopic,  the pointer observable 
also obeys [17]
$[A,H]=0$, with $H$ the system-environment interaction Hamiltonian.  
Let us now write the density matrix evolution in Schr\"odinger picture,  
$$d\rho/dt =-i[H_{\rm TOT}, \rho]= -i[H_{\cal S}+H_{\cal E}+H,\rho]
~~~,\eqno(60a)$$
with $H_{\cal S}$ the system Hamiltonian.  We consider the system evolution 
after a brief interaction has entangled the apparatus states with 
the microscopic subsystem quantum states that are to be distinguished by the pointer 
reading.   For the time evolution of the 
variance of the pointer observable  $A$, we have 
$$dV/dt = {\rm Tr} (d\rho/dt) A^2-2 ({\rm Tr} \rho A)({\rm Tr} (d\rho/dt) A)  
~~~,\eqno(60b)$$
which substituting Eq.~(60a), and using the cyclic property of the trace 
and the fact that $A$ commutes with both $H_{\cal E}$ and $H$, simplifies 
to 
$$dV/dt= i {\rm Tr} \rho [H_{\cal S}, A^2]  
-2i ({\rm Tr}\rho A) ({\rm Tr} \rho [H_{\cal S}, A])
~~~.\eqno(60c)$$ 
This can be further simplified by using the definition of the 
reduced density matrix $\rho^{(1)}={\rm Tr}_{\cal E} \rho$, 
together with  the fact that the commutators in Eq.~(60c) involve 
only system operators, giving  
$$dV/dt= i {\rm Tr}_{\cal S} \rho^{(1)}  [H_{\cal S}, A^2]  
- 2i ({\rm Tr}_{\cal S}\rho^{(1)} A)   
({\rm Tr}_{\cal S}\rho^{(1)} [H_{\cal S}, A] )
~~~.\eqno(60d)$$ 
We see that, unlike the It\^o equation case discussed above, the time 
derivative of $V$ here is determined by $\rho^{(1)}$, rather than by 
$\rho^{(2)}$.  

Let us now take the pointer observable to be a pointer center of mass 
coordinate $A=X$, in which case, once the entanglement of the pointer 
with the microsystem being measured has been established, 
 the relevant part of the system Hamiltonian 
$H_{\cal S}$ is $P^2/(2M)$, with $P$ the total momentum operator for the 
pointer of macroscopic mass $M$.  Evaluating the commutators, and 
writing ${\rm Tr}_{\cal S}  \rho^{(1)} {\cal O}=\langle {\cal O}\rangle$, 
we see that 
$$
\eqalign{
dV/dt =&  (1/M) \langle \{P,X\} \rangle 
-(2/M)\langle X \rangle  \langle P\rangle\cr
=&(1/M) \langle \{ X-\langle X\rangle,P-\langle P\rangle \}\rangle   
~~~.\cr
}\eqno(61a)$$
By the Schwartz inequality, the right hand 
side of Eq.~(61a) is bounded by 
$$(2/M) \langle (X-\langle X\rangle)^2 \rangle^{1\over 2} 
 \langle (P-\langle P\rangle)^2 \rangle^{1\over 2}  
 =(2/M)\Delta X \Delta P~~~.
\eqno(61b)$$

Let us now determine the minimum value of the bound of Eq.~(61b) that is 
compatible with the parameters of a feasible measurement. Since  
 the uncertainty principle implies that  $\Delta X 
\Delta P \geq 1/2$,  we get a least 
upper bound on  Eq.~(61b) by 
substituting  $\Delta X \Delta P \sim 1/2$. This shows that  
 $|dV/dt|$ can  be made as small as $\sim 1/M$,    
which since $M$ is macroscopic, can be made essentially arbitrarily 
small.\footnote{$^3$}{Restoring factors of Planck's constant, $|dV/dt|$ can  
be as small as $\hbar/M$, for which the reduction time $dt$ is at least 
of order $MdV/\hbar$.  For $M \sim 10^{24} m_{\rm proton}$ and 
$dV \sim (1 {\rm cm})^2$, this gives $dt \sim M (1 {\rm cm})^2/\hbar 
\sim 10^{27} {\rm s} \sim 10^{10}$ times the age of the universe.  Note 
that our argument places no restriction on the mean pointer momentum 
$\langle P \rangle$ that establishes the time needed to attain one or 
the other of the measurement outcomes $X$ starting from the initial 
pointer position.}  
Hence the variance of the pointer variable $A$ stays essentially 
constant, and is not forced to reduce to zero in the course of the 
measurement.  

We conclude, in agreement with the arguments of [16],  that a quantum apparatus interacting with a 
quantum environment does not  act like the  stochastic equation 
of Eq.~(57a) in terms of reducing the state vector. 
Although a quantum environment acts on 
a quantum system with a form of ``noise'', our analysis of the density 
tensor hierarchy in the classical and quantum noise cases shows that 
structures with  different kinematical symmetries,\footnote{$^4$}{The 
dissimilarities between the symmetries of the classical 
noise and quantum noise hierarchies are least for the order two density 
tensor.   
In the order two case, cyclic symmetry is equivalent to full permutation 
symmetry, and so the index symmetry properties are the same in the classical 
and quantum noise cases, 
and as a consequence the descent equations in the quantum noise case 
correspond to  the 
idempotence descent equations in the classical noise case. Only the 
classical 
noise descent equation implied by the unit trace condition has no precise  quantum noise counterpart: in the classical case, one has   
$$\delta_{i_1j_1} \rho^{(2)}_{i_1j_1,i_2j_2}= 
\delta_{i_1j_1} E[\rho_{i_1j_1}\rho_{i_2j_2}]=E[\rho_{i_2j_2}]
=\rho^{(1)}_{i_2j_2}~~~$$
whereas in the quantum noise case one instead has 
$$\delta_{i_1j_1} \rho^{(2)}_{i_1j_1,i_2j_2}= 
\delta_{i_1j_1} {\rm Tr}_{\cal E}\rho_{i_1j_1}\rho_{i_2j_2} 
= {\rm Tr}_{\cal E} \rho_{\cal E} \rho_{i_2j_2}   
\not = {\rm Tr}_{\cal E}  \rho_{i_2j_2}  =\rho^{(1)}_{i_2j_2} 
~~~,$$ 
with $\rho_{\cal E}={\rm Tr}_{\cal S} \rho$ the reduced density matrix of the environment with the system traced out.}
different  dynamical evolutions,  
and different implications for the measurement process are involved.   As a result, 
the quantum noise in a closed 
quantum system does not mimic the action of the classical noise in objective 
reduction models, and cannot be invoked to give a resolution of the    
quantum measurement problem within the framework of unmodified 
quantum mechanics.
\bigskip

\centerline{\bf Acknowledgments}

I wish to thank Angelo Bassi, Todd Brun, 
Lajos Di\'osi, Larry Horwitz, and Lane Hughston for 
instructive conversations over a number of years that helped motivate 
this study, and Francesco Petruccione for the gift some years ago 
of a copy of ref [1].  
This work was supported in part by the Department of Energy under
Grant \#DE--FG02--90ER40542.
\bigskip

\centerline{\bf Appendix A: Descent equations for the}  
\centerline{\bf isotropic spin-1/2 ensemble}
\bigskip
Let us write the generating function of Eq.~(14b) as 
$$\eqalign{
G[a_{ij}]=&fg~~~,\cr
f(x)=&\sinh x^{1\over 2}/x^{1\over 2}~,~~x=\vec A^{\,2}~~~,\cr
g=&e^{{1\over 2} {\rm Tr}a}~~~.\cr
}\eqno(A1)$$
Then, we find 
$$\eqalign{
{\partial G \over \partial a_{mr}}
=&{1\over 2} \delta_{mr} G + g f^{\prime} 
\vec A \cdot \vec \sigma_{mr}~~~,\cr
{\partial^2 G \over \partial a_{mr} \partial a_{pq} } =&
{1\over 2}\delta_{mr} {\partial G \over \partial a_{pq}} 
+ {1\over 2} \delta _{pq} g f^{\prime} \vec A \cdot \vec \sigma_{mr} \cr
+&g f^{\prime \prime} 
 \vec A \cdot \vec \sigma_{mr} \vec A \cdot \vec \sigma_{pq}  
+ {1\over 2} g f^{\prime} \vec \sigma_{mr} \cdot \vec \sigma_{pq}~~~.\cr
}\eqno(A2)$$ 
Here the primes denote derivatives of $f$ with respect to $x$, and in this 
notation $f$ obeys the second order differential equation 
$$xf^{\prime \prime} + {3\over 2} f^{\prime} ={1\over 4} f~~~.\eqno(A3)$$
Contracting the first expression in  Eq.~(A2) with $\delta_{mr}$, and 
using the tracelessness of the Pauli matrices, gives the first equation 
in Eq.~(7b).  Contracting the second expression in Eq.~(A2) with 
$\delta_{rp}$, and using the differential equation of Eq.~(A3) together 
with the Pauli matrix identities $(\vec \sigma^{\,2})_{mq}=3 \delta_{mq}$ 
and $\sigma^i\sigma^j=\delta^{ij} + i \epsilon^{ijk} \sigma^k$, which 
implies $ \vec A \cdot \vec \sigma_{mp} \vec A \cdot \vec \sigma_{pq}   
=\vec A^{\,2} \delta_{mq}$, gives the second equation in Eq.~(7b).  
\bigskip

\centerline{\bf Appendix B: Descent equations for the}  
\centerline{\bf It\^o stochastic Schr\"odinger equation}  
\bigskip
We wish here to verify that 
$$dG[a]=dt E\big[\big(a_{mr}({\cal L}\rho)_{mr}           
+{1\over 2}a_{mr}a_{pq} C_{mr,pq}\big)e^{\rho \cdot a}\big]    
\eqno(B1)$$   
obeys the descent equations of Eq.~(7b).  Since 
$$\delta_{mr} ({\cal L} \rho)_{mr}=\delta_{mr} C_{mr,pq} = 
\delta_{pq} C_{mr,pq}=0~~~,\eqno(B2a)$$ 
we have 
$$\delta_{uv}{\partial dG[a] \over \partial a_{uv} }=
dt E\big[\big(a_{mr}({\cal L}\rho)_{mr}           
+{1\over 2}a_{mr}a_{pq} C_{mr,pq}\big)({\rm Tr}\rho) 
e^{\rho\cdot a}\big]= dG[a] ~~~, \eqno(B2b)$$   
giving the first identity in Eq.~(7b).  Next we calculate 
$$\eqalign{
{\partial dG[a] \over \partial a_{mq} }=&dt 
E\big[\big(({\cal L}\rho)_{mq}           
+{1\over 2}a_{uv} (C_{mq,uv}+C_{uv,mq})\big) e^{\rho\cdot a}  \cr
+&\big(a_{uv} ({\cal L}\rho)_{uv}   
+{1\over 2}a_{uv}a_{rs} C_{uv,rs}\big) \rho_{mq}
  e^{\rho\cdot a}\big]   ~~~,\cr 
}\eqno(B3a)$$
while for the contraction of the second variation we have (with indices $m,q$ implicit on 
the right hand side) 
$$
{\partial^2  dG[a] \over \partial a_{mr} \partial a_{rq} }= 
dt E\big[\big(S_1+S_2+a_{uv}(T_{1uv} +T_{2uv})\big) e^{\rho\cdot a} \big]~~~,\eqno(B3b)$$
with 
$$\eqalign{ 
S_1=&{1\over 2} (C_{mr,rq} + C_{rq,mr})~~~,\cr
S_2=&\{ {\cal L}\rho, \rho\}_{mq} ~~~,\cr
T_{1uv}=&{1\over 2}[ (C_{mr,uv}+C_{uv,mr})\rho_{rq}
+ \rho_{mr} (C_{uv,rq}+C_{rq,uv}) ]~~~,\cr
T_{2uv}=& \big[({\cal L} \rho)_{uv}+{1\over 2} a_{rs}C_{uv,rs}\big] \rho_{mq}~~~,\cr
}\eqno(B3c)$$
We see immediately that $a_{uv}T_{2uv}$  gives all 
of the second line of Eq.~(B3a).  
From Eqs.~(16b) and (18) we find 
$$\{ {\cal L}\rho, \rho\}_{mq} 
= ({\cal L}\rho)_{mq}
-[(c_k-\langle c_k \rangle) \rho 
(c_k-\langle c_k \rangle)^{\dagger} ]_{mq} -\rho_{mq} 
\langle  (c_k-\langle c_k \rangle)^{\dagger} (c_k-\langle c_k \rangle) \rangle
~~~,\eqno(B4a)$$
while from Eq.~(21b) we have 
$$ {1\over 2} (C_{mr,rq} + C_{rq,mr})
=[(c_k-\langle c_k \rangle) \rho 
(c_k-\langle c_k \rangle)^{\dagger} ]_{mq} +\rho_{mq} 
\langle  (c_k-\langle c_k \rangle)^{\dagger} (c_k-\langle c_k \rangle) \rangle
~~~.\eqno(B4b)$$ 
Hence $S_1+S_2=({\cal L}\rho)_{mq}$, giving the ${\cal  L} \rho$ part of the 
first line of Eq.~(B3a).   Finally, again using Eq.~(21b) we find 
that 
$$
{1\over 2}[ (C_{mr,uv}+C_{uv,mr})\rho_{rq}
+ \rho_{mr} (C_{uv,rq}+C_{rq,uv}) ]={1\over 2}
(C_{mq,uv}+C_{uv,mq})~~~,\eqno(B4c)$$ 
and so $a_{uv}T_{1uv}$ gives the remainder of the first line of Eq.~(B3a), 
completing the check of the descent equations.  
\bigskip

\centerline{\bf Appendix C: Descent equations for the} 
\centerline{\bf jump Schr\"odinger equation}
\bigskip
We verify here that 
$$dG[a]= 
dt E\big[(a \cdot {\cal L} \rho + \sum_{p=2}^{\infty} \sum_k v_k     
{(a \cdot Q_k)^p \over p!} ) e^{a \cdot \rho}\big] ~~~\eqno(C1a)$$
obeys the descent equations of Eq.~(7b).  Since  ${\rm Tr}( {\cal L}\rho)=0$ 
and ${\rm Tr} Q_k=\langle B_k+B_k^{\dagger}+B_k^{\dagger}B_k\rangle=0$, 
we have 
$$ \delta_{uv}{\partial dG[a] \over \partial a_{uv} }=
dt E\big[(a \cdot {\cal L} \rho + \sum_{p=2}^{\infty} \sum_k v_k 
{(a \cdot Q_k)^p \over p!} )({\rm Tr} \rho) e^{a \cdot \rho}\big] 
~~~,\eqno(C1b)$$
checking the first line of Eq.~(7b).   Next we calculate the first 
variation of $G$,
$$\eqalign{
{\partial dG[a] \over \partial a_{mq} }=&dt 
E\big[\big(({\cal L}\rho)_{mq} +   \sum_{p=2}^{\infty} \sum_k v_k         
{(a \cdot Q_k)^{p-1} \over (p-1)!  } (Q_k)_{mq} \big)  e^{a \cdot \rho}\cr
+& (a \cdot {\cal L} \rho + \sum_{p=2}^{\infty} \sum_k v_k     
{(a \cdot Q_k)^p \over p!} )\rho_{mq} e^{a \cdot \rho}\big]~~~,  \cr
}\eqno(C2a)$$
and the contracted second variation, 
$$
{\partial^2  dG[a] \over \partial a_{mr} \partial a_{rq} }= 
dt E\big[(S_1+S_2+S_3+S_4) e^{a \cdot \rho} \big]~~~,\eqno(C2b)$$
with 
$$\eqalign{
S_1=&  \sum_{p=2}^{\infty} \sum_k v_k  {(a \cdot Q_k)^{p-2} \over (p-2)!}   
 (Q_k^2)_{mq} ~~~,\cr
S_2=& \{ {\cal L}\rho, \rho\}_{mq} ~~~,\cr
S_3=&  \sum_{p=2}^{\infty} \sum_k v_k {(a \cdot Q_k)^{p-1} \over (p-1)!  }   
\{Q_k,\rho\}_{mq}~~~,\cr
S_4=&\left(a \cdot {\cal L} \rho + \sum_{p=2}^{\infty} \sum_k v_k     
{(a \cdot Q_k)^p \over p!} \right)\rho_{mq}~~~.  \cr
}\eqno(C2c)$$
We see immediately that $S_4$ gives all of the second line of 
Eq.~(C2a).   From Eq.~(30c), which we rewrite here,  
$$\eqalign{
\{{\cal L}\rho,\rho\}=&{\cal L}\rho-\sum_k v_k Q_k^2~~~,\cr
\{Q_k,\rho\}=&Q_k-Q_k^2~~~, \cr
}\eqno(C3a)$$ 
we see that the  ${\cal L}\rho$ part of $S_2$  and the $Q_k$ part 
of $\{Q_k,\rho\}$  in $S_3$ 
give the first line of Eq.~(C2a).  To complete the verification, we 
must show that $S_1$ cancels against the remainder of $S_2+S_3$, which is   
$$-\sum_k v_k Q_k^2 
-  \sum_{p=2}^{\infty} \sum_k v_k {(a \cdot Q_k)^{p-1} \over (p-1)!  } 
 (Q_k^2)_{mq}~~~.\eqno(C3b)$$ 
But separating off the $p=2$ term of $S_1$, and making the change 
of variable $p\to p+1$ in the remaining sum, we see that $S_1$ is exactly 
the negative of Eq.~(C3b), completing the argument.  
\bigskip

\centerline{\bf Appendix D: Descent equations for the Born-Markov} 
\centerline{\bf master equation}
\bigskip
We wish here to verify that Eq.~(50a) obeys the descent equations 
of Eq.~(34).  We separate the verification into two parts, first 
checking the descent from $n=2$ to $n=1$, and then checking the 
descent from general $n > 2$ to $n-1$.  For the $n=2$ density tensor 
time derivative, writing out all terms in Eq.~(50a) explicitly,  
and using the fact that since operators labeled with subscripts $2$ and $1$ 
act on different Hilbert spaces, the order in which they are 
written is irrelevant, we have 
$$\eqalign{
d\rho^{(2)}(t)/dt 
=&  i[\rho^{(1)}_1(t), \sum_{\omega \alpha \beta} 
S_{\alpha \beta}(\omega) A_{1 \alpha}^{\dagger}(\omega) 
A_{1 \beta}(\omega)]   \rho^{(1)}_2(t)  \cr
+&   \rho^{(1)}_1(t) 
i[\rho^{(1)}_2(t), \sum_{\omega \alpha \beta} 
S_{\alpha \beta}(\omega) A_{2 \alpha}^{\dagger}(\omega) 
A_{2 \beta}(\omega)] \cr
-&\sum_{\omega\alpha\beta}\gamma_{\alpha \beta}(\omega) 
{1\over 2}\{ A_{1 \alpha}^{\dagger} (\omega)A_{1 \beta}(\omega),  
\rho^{(1)}_1(t) \}  \rho^{(1)}_2(t)   \cr 
-&\sum_{\omega\alpha\beta}\gamma_{\alpha \beta}(\omega)  
\rho^{(1)}_1(t)
{1\over 2}\{ A_{2 \alpha}^{\dagger} (\omega)A_{2 \beta}(\omega),  
\rho^{(1)}_2(t) \}  \cr 
+&\sum_{\omega\alpha\beta}\gamma_{\alpha \beta}(\omega)    
[\rho^{(1)}_1(t) A_{1 \alpha}^{\dagger}(\omega) 
A_{2\beta}(\omega)\rho^{(1)}_2(t)  
+A_{1\beta}(\omega)\rho^{(1)}_1(t) 
\rho^{(1)}_2(t) A_{2 \alpha}^{\dagger}(\omega) ]
~~~. }\eqno(D1a)$$

Contracting the column index associated with the subscript 1 with the 
row index associated with the subscript 2, and dropping the subscripts  
since all operators now act in the same Hilbert space, we get 
$$\eqalign{
d\rho^{(2)}(t)/dt 
\to &  i[\rho^{(1)}(t), \sum_{\omega \alpha \beta} 
S_{\alpha \beta}(\omega) A_{ \alpha}^{\dagger}(\omega) 
A_{ \beta}(\omega)]   \rho^{(1)}(t)  \cr
+&   \rho^{(1)}(t) 
i[\rho^{(1)}(t), \sum_{\omega \alpha \beta} 
S_{\alpha \beta}(\omega) A_{ \alpha}^{\dagger}(\omega) 
A_{ \beta}(\omega)] \cr
-&\sum_{\omega\alpha\beta}\gamma_{\alpha \beta}(\omega) 
{1\over 2}\{ A_{ \alpha}^{\dagger} (\omega)A_{ \beta}(\omega),  
\rho^{(1)}(t) \}  \rho^{(1)}(t)   \cr 
-&\sum_{\omega\alpha\beta}\gamma_{\alpha \beta}(\omega)        
\rho^{(1)}(t)
{1\over 2}\{ A_{ \alpha}^{\dagger} (\omega)A_{ \beta}(\omega),  
\rho^{(1)}(t) \}  \cr 
+&\sum_{\omega\alpha\beta}\gamma_{\alpha \beta}(\omega)    
[\rho^{(1)}(t) A_{ \alpha}^{\dagger}(\omega) 
A_{\beta}(\omega)\rho^{(1)}(t)  
+A_{\beta}(\omega)(\rho^{(1)}(t))^2   A_{ \alpha}^{\dagger}(\omega) ] \cr
=&i[(\rho^{(1)}(t) )^2, \sum_{\omega \alpha \beta} 
S_{\alpha \beta}(\omega) A_{ \alpha}^{\dagger}(\omega) 
A_{ \beta}(\omega)]    \cr
+&  \sum_{\omega\alpha\beta}\gamma_{\alpha \beta}(\omega)        
\left[ A_{\beta}(\omega)(\rho^{(1)}(t))^2   A_{ \alpha}^{\dagger}(\omega) 
-{1\over 2} \{ (\rho^{(1)}(t) )^2,  A_{ \alpha}^{\dagger}(\omega) 
A_{ \beta}(\omega)  \}\right]
~~~, \cr  
}\eqno(D1b)$$
which has the structure of $d\rho^{(1)}(t)/dt$ and so verifies the  
$2 \to 1$ descent.  

To verify the $n \to n-1$ descent we make some simplifications in notation. 
We omit all superscripts $(1)$, since this leads to no ambiguities, as well as  
all time arguments $(t)$ and all frequency arguments $(\omega)$.
We also abbreviate 
$$\eqalign{
L_{\ell}\equiv&
 \sum_{\omega \alpha \beta} 
S_{\alpha \beta}(\omega) 
A_{\ell \alpha}^{\dagger}(\omega) A_{\ell \beta}(\omega)~~~,    \cr
M_{\ell}\equiv&\sum_{\omega\alpha\beta}\gamma_{\alpha \beta}(\omega)    
A_{\ell \alpha}^{\dagger}(\omega) A_{\ell \beta}(\omega)~~~.\cr 
}\eqno(D2a)$$
Our general strategy is to split the sum  $\sum_{\ell=1}^n$  
containing $(\rho_1...\rho_n)_{\ell}$ into $\sum_{\ell=2}^{n-1}$ plus 
the $\ell=1$ and the $\ell=n$ terms, and to split the sum   $\sum_{\ell=1}^n$ 
containing  $(\rho_1...\rho_n)_{\ell\ell+1}$ into $\sum_{\ell=2}^{n-2}$ plus 
the  $\ell=1$, $\ell=n-1$, and  $\ell=n$  terms.  For the part of 
$d\rho^{(n)}/dt$ 
involving $L_{\ell}$, we have 
$$\sum_{\ell=2}^{n-1} (\rho_1...\rho_{n-1})_{\ell} \rho_n i 
[\rho_{\ell},L_{\ell}]  +(\rho_2...\rho_n) i[\rho_1,L_1] 
+(\rho_1...\rho_{n-1}) i [\rho_n,L_n]~~~,\eqno(D2b)$$
which on contracting the column index associated with the subscript 
$n$ with the row index associated with the subscript $1$, and relabeling  
all quantities that had subscript $n$ with subscript $1$, since they act now 
in the same Hilbert space, gives 
$$\eqalign{
&\sum_{\ell=2}^{n-1}(\rho_1^2\rho_2...\rho_{n-1})_{\ell}
i[\rho_{\ell},L_{\ell}]  
+\rho_2...\rho_{n-1}  i(\rho_1 [\rho_1,L_1]+ [\rho_1,L_1]\rho_1)    \cr   
=&\sum_{\ell=2}^{n-1}(\rho_1^2\rho_2...\rho_{n-1})_{\ell}i[\rho_{\ell},L_{\ell}]   
+\rho_2...\rho_{n-1}  i [\rho_1^2,L_1] ~~~,    \cr   
}\eqno(D2c)$$
which has the correct structure for the corresponding part of 
$d\rho^{(n-1)}/dt$, with $\rho_1$ replaced by $\rho_1^2$.  
The remainder of $d\rho^{(n)}/dt$ is 
$$\eqalign{
&-\sum_{\ell=2}^{n-1}(\rho_1...\rho_n)_{\ell} {1\over 2} 
\{M_{\ell},\rho_{\ell} \} 
-(\rho_2...\rho_n) {1\over 2} \{ M_1,\rho_1\}    
-(\rho_1...\rho_{n-1}) {1\over 2} \{ M_n,\rho_n\} \cr   
+& \sum_{\omega\alpha\beta}\gamma_{\alpha \beta} \left(   
\sum_{\ell=2}^{n-2} (\rho_1...\rho_n)_{\ell \ell+1} \rho_{\ell} 
A_{\ell \alpha}^{\dagger}  A_{\ell+1 \beta} \rho_{\ell+1} 
+\rho_3...\rho_n \rho_1 A_{1\alpha}^{\dagger} 
A_{2 \beta} \rho_2 \right.\cr
+&\left. \rho_1...\rho_{n-2}   \rho_{n-1}A_{n-1 \alpha}^{\dagger}
A_{n\beta} \rho_n
+\rho_2...\rho_{n-1}\rho_n A_{n\alpha}^{\dagger} A_{1\beta} \rho_1 \right)~~~.\cr
}\eqno(D3a)$$
Again,  contracting the column index associated with the subscript 
$n$ with the row index associated with the subscript $1$,   and relabeling  
all quantities that had subscript $n$ with subscript $1$, since they act now 
in the same Hilbert space, gives 
$$\eqalign{ 
&-\sum_{\ell=2}^{n-1}(\rho_1^2...\rho_{n-1})_{\ell} {1\over 2} 
\{M_{\ell},\rho_{\ell} \} 
-(\rho_2...\rho_{n-1}) \left( \rho_1 M_1\rho_1 (*)   
+{1\over 2} \{ M_1,\rho_1^2\} \right)  \cr   
+& \sum_{\omega\alpha\beta}\gamma_{\alpha \beta} \left(   
\sum_{\ell=2}^{n-2} (\rho_1^2...\rho_{n-1})_{\ell \ell+1} \rho_{\ell} 
A_{\ell \alpha}^{\dagger}  A_{\ell+1 \beta} \rho_{\ell+1} 
+(\rho_3...\rho_{n-1}) \rho_1^2 A_{1\alpha}^{\dagger} 
A_{2 \beta} \rho_2 \right.\cr
+&\left. \rho_2...\rho_{n-2} \rho_{n-1}A_{n-1 \alpha}^{\dagger}
A_{1\beta} \rho_1^2
+\rho_2...\rho_{n-1} \rho_1 A_{1\alpha}^{\dagger} A_{1\beta} \rho_1  (*) 
\right)~~~,\cr
}\eqno(D3b)$$
which on canceling the terms marked with $(*)$ gives the corresponding part 
of $d\rho^{(n-1)}/dt$, with $\rho_1$ replaced by $\rho_1^2$.   This 
completes the verification of the $n \to n-1$ descent. 
\bigskip

\centerline{\bf Appendix E: Descent equations for the} 
\centerline{\bf Caldeira--Leggett model}
\bigskip
We verify here that Eqs.~(56a) and (56b) obey the descent equations 
of Eq.~(34).  As in the preceding appendix, we simplify the notation 
by omitting all superscripts $(1)$ and all time arguments $(t)$.  We 
first verify the $n=2$ to $n=1$ descent.  For the $n=2$ case of Eq.~(56b), 
we have 
$$\eqalign{
D\rho^{(2)}/dt=
& \rho_2 
 [ -2m\gamma k_B T  (x_1^2 \rho_1 +\rho_1 x_1^2) 
 + i\gamma(\rho_1p_1x_1-x_1p_1\rho_1) ]  \cr 
+& \rho_1 
 [ -2m\gamma k_B T  (x_2^2 \rho_2 +\rho_2 x_2^2) 
 + i\gamma(\rho_2p_2x_2-x_2p_2\rho_2) ]  \cr 
+& 4m\gamma k_B T (\rho_1 x_1 x_2 \rho_2 + \rho_2 x_2 x_1\rho_1) 
 +i\gamma[ \rho_1 (x_1 p_2-p_1 x_2) \rho_2+ \rho_2 (x_2 p_1-p_2 x_1) \rho_1]    
 ~~~. \cr
 }\eqno(E1A)$$
Contracting the column index associated with the subscript 1 with the 
row index associated with the subscript 2, and dropping subscripts since 
all operators now act in the same Hilbert space, we get 
$$\eqalign{
D\rho^{(2)}/dt\to 
& -2m\gamma k_B T  (x^2 \rho^2 +\rho x^2 \rho) 
 + i\gamma(\rho p x \rho -x p \rho^2)   \cr 
& -2m\gamma k_B T  (\rho x^2 \rho +\rho^2 x^2) 
 + i\gamma(\rho^2  p x -\rho x p \rho)   \cr 
+& 4m\gamma k_B T (\rho x^2 \rho  + x \rho^2 x)
 +i\gamma[ \rho (x p -p x) \rho+  p \rho^2 x - x \rho^2 p ]    
 ~~~. \cr
 }\eqno(E1B)$$
We see that the terms that have an operator sandwiched between two factors 
of $\rho$ cancel, leaving only terms involving $\rho^2$, which have the 
form of Eq.~(56a) with $\rho$ replaced by $\rho^2$.  

To check the $n>2$ to $n-1$ descent, we split the sums that occur in the 
same manner as in Appendix D.  
We thus write Eq.~(56b) in the form 
$$\eqalign{
D\rho^{(n)}/dt=& 
\sum_{\ell=2}^{n-1} 
(\rho_1...\rho_n)_{\ell} [-2m\gamma k_B T \{x_{\ell}^2,\rho_{\ell} \} 
+ i\gamma(\rho_{\ell}p_{\ell}x_{\ell}-x_{\ell}p_{\ell}\rho_{\ell} )] \cr
+&\rho_2...\rho_n [-2m\gamma k_B T \{x_1^2,\rho_1 \} 
+ i\gamma(\rho_1p_1x_1-x_1p_1\rho_1 )] \cr
+&\rho_1...\rho_{n-1} [-2m\gamma k_B T \{x_n^2,\rho_n \} 
+ i\gamma(\rho_n p_nx_n-x_np_n\rho_n )] \cr
+&\sum_{\ell=2}^{n-2}(\rho_1...\rho_n)_{\ell \ell+1} 
[4m\gamma k_B T \rho_{\ell} x_{\ell} x_{\ell +1}  \rho_{\ell+1} 
+i\gamma \rho_{\ell} (x_{\ell}p_{\ell+1}-p_{\ell}x_{\ell+1}) \rho_{\ell+1}]\cr
+&\rho_3...\rho_n [4m\gamma k_B T \rho_1 x_1 x_2  \rho_2 
+i\gamma \rho_1 (x_1p_2-p_1x_2) \rho_2 ]\cr
+&\rho_1...\rho_{n-2} [4m\gamma k_B T \rho_{n-1} x_{n-1} x_n  \rho_n 
+i\gamma \rho_{n-1} (x_{n-1}p_n-p_{n-1}x_n) \rho_n ]\cr
+&\rho_2...\rho_{n-1} [4m\gamma k_B T \rho_n x_n x_1  \rho_1 
+i\gamma \rho_n (x_np_1-p_nx_1) \rho_1 ]~~~.\cr
}\eqno(E2)$$
We now contract the column index associated with the subscript 
$n$ with the row index associated with the subscript $1$,   and relabel   
all quantities that had subscript $n$ with subscript $1$, since they act now 
in the same Hilbert space.  As is readily seen by inspection of Eq.~(E2), 
this gives Eq.~(56b) with $n$ replaced by $n-1$ and 
with $\rho_1$ replaced by $\rho_1^2$, together with  terms of the wrong 
structure, that grouped  
together give 
$(4-2-2) \rho_2...\rho_{n-1} m\gamma k_B T \rho_1 x_1^2 \rho_1=0$  and 
$(1-1)\rho_2...\rho_{n-1} i\gamma \rho_1 (x_1p_1-p_1x_1) \rho_1=0$,  
 which thus vanish.   This completes the verification of the descent 
 equation for Eq.~(56b).  

\bigskip

\vfill\eject
\centerline{\bf References}
\bigskip
\noindent
[1] Breuer H-P and Petruccione F (2002) {\it The Theory of Open 
Quantum Systems} (Oxford: Oxford University Press)      \hfill\break
\bigskip 
\noindent
[2] Breuer H-P and Petruccione F (1996) {\it Phys. Rev. A} {\bf 54} 1146 
\hfill\break
\bigskip
\noindent
[3] Wiseman H M (1993) {\it Phys. Rev. A} {\bf 47} 5180  \hfill\break 
\bigskip
\noindent
[4] M\o lmer K, Castin Y and Dalibard J (1993) {\it J. Opt. Soc. Am. B}  
{\bf 10} 524 \hfill\break
\bigskip
\noindent
[5] Mielnik, B (1974) {\it Commun. Math. Phys.} {\bf 37} 221; see 
especially p 240.  I wish to thank Lane Hughston for bringing this 
reference, and ref [6] as well, to my attention.   \hfill\break
\bigskip
\noindent
[6] Brody, D C and Hughston, L P (1999) {\it J.  Math. Phys.} {\bf 40} 
12, Eqs.~(31) and (32); Brody, D C and Hughston, L P (1999) 
{\it Proc. Roy. Soc.  A} {\bf 455} 1683, Sec. 2(e); 
 Brody, D C and Hughston, L P (2000) {\it J. Math. Phys.} 
{\bf 41}, 2586, Eq.~(9) and subsequent discussion. \hfill\break
\bigskip
\noindent
[7] Wiseman H M and Di\'osi L (2001)  {\it Chem. Phys.} {\bf 268} 91.  
See also Di\'osi L (1986) {\it Phys. Lett. A} 
{\bf 114} 451 for the transition rate operator.  \hfill\break 
\bigskip
\noindent
[8] Lindblad G (1976) {\it Commun. Math. Phys.} {\bf 48} 119 
\hfill\break
\bigskip
\noindent
[9] Gorini V, Kossakowski A and Sudarshan E C G (1976) {\it J. Math. 
Phys.} {\bf 17} 821  \hfill\break
\bigskip
\noindent
[10] Schack R and Brun T A (1997) {\it Comp. Phys. Commun.} {\bf 102} 210 
\hfill\break 
\bigskip
\noindent
[11] Gallis M R and Fleming G N (1990) {\it Phys. Rev. A} {\bf 42} 38
\hfill\break
\bigskip
\noindent
[12] Di\'osi L (1995) {\it Europhys. Lett.} {\bf 30} 63; Dodd P J and 
Halliwell J J (2003) {\it Phys. Rev. D} {\bf 67} 105018; Hornberger K 
and Sipe J E (2003) {\it Phys. Rev. A} {\bf 68} 012105; Adler S L 
(2006) {\it J. Phys. A: Math. Gen.} {\bf 39} 14067 \hfill\break 
\bigskip
\noindent
[13] Hornberger K (2006)  Introduction to decoherence theory,  
arXiv: quant-ph/0612118 \hfill\break
\bigskip
\noindent
[14]  Caldeira A O and Leggett A J (1983) {\it Physica A} {\bf 121} 
587  \hfill\break
\bigskip
\noindent
[15]  Ghirardi G C, Pearle P and Rimini A (1990) {\it Phys. Rev. A} 
{\bf 42} 78; Hughston L P (1996) {\it Proc. Roy. Soc.  A} 
{\bf 452} 953; Adler S L and Horwitz L P (2000) {\it J. Math. Phys.} 
{\bf 41} 2485; Adler S L, Brody D C, Brun T A and Hughston L P (2001)
{\it J Phys. A: Math. Gen.} {\bf 34} 8795; Adler S L (2004) {\it Quantum 
Theory as an Emergent Phenomenon} (Cambridge UK: Cambridge University Press) Sec. 6.2\hfill\break 
\bigskip
\noindent
[16] Bassi A and Ghirardi G C {\it Phys. Lett. A} {\bf 275} 373 
\hfill\break
\bigskip
\noindent
[17] Zurek W H (1981) {\it Phys. Rev. D} {\bf 24} 1516; 
Schlosshauer  M (2004) {\it Rev. Mod. Phys.} {\bf 75} 1267, p. 1280
 \hfill\break
\bigskip
\noindent
[18]  For reviews of stochastic reduction models, see Bassi A and 
Ghirardi G C (2003) {\it Phys. Reports} {\bf 379} 257; Pearle P (1999) 
Collapse models, in {\it Open Systems and Measurements in Relativistic 
Quantum Field Theory}, Lecture Notes in Physics 526, Breuer H-P and 
Petruccione F eds. (Berlin: Springer-Verlag) 
\hfill\break 
\bigskip
\noindent
\bigskip
\noindent
\bigskip
\noindent
\bigskip
\noindent
\bigskip
\noindent
\vfill
\eject
\bigskip
\bye